\documentclass[a4paper,11pt]{article}
\pdfoutput=1 

\usepackage{jcappub} 

\usepackage[T1]{fontenc} 
\usepackage{subfig}
\usepackage{multirow}
\usepackage{verbatim}
\usepackage{comment}
\usepackage[table]{xcolor}

\title{\boldmath Constraints on the Spacetime Dynamics of an Early Dark Energy Component}

\author[a,b,c]{Hasti Khoraminezhad \note{email: hkhorami@sissa.it}}
\author[a,b,c,d]{Matteo Viel}
\author[a,b,c]{Carlo Baccigalupi}
\author[e,f]{Maria Archidiacono}

\affiliation[a]{SISSA, International School for Advanced Studies, Via Bonomea 265, 34136 Trieste, Italy}
\affiliation[b]{INFN, Sezione di Trieste, Via Valerio 2, 34127 Trieste, Italy}
\affiliation[c]{IFPU, Institute for Fundamental Physics of the Universe, Via Beirut 2, 34014 Trieste, Italy}
\affiliation[d]{INAF/OATS, Osservatorio Astronomico di Trieste, Via G.B. Tiepolo 11, 34143 Trieste, Italy}
\affiliation[e]{Universit\`a degli Studi di Milano, via G. Celoria 16, 20133 Milano, Italy}
\affiliation[f]{INFN, Sezione di Milano, Milano, Italy}

\abstract{
We consider an Early Dark Energy (EDE) cosmological model, and perform an analysis which takes into account both background and perturbation effects via the parameters $c^{2}_{\rm eff}$ and $c^{2}_{\rm vis}$, representing effective sound speed and viscosity, respectively. 
By using the latest available data we derive constraints on the amount of dark energy at early times and the present value of the equation of state. Our focus is on the effect that early dark energy has on the Cosmic Microwave Background (CMB) data, including polarization and lensing, in a generalized parameter space including a varying total neutrino mass, and tensor to scalar ratio, besides the 6 standard parameters of the minimal cosmological model. We find that the inclusion of Baryonic Acoustic Oscillations (BAO) data and CMB lensing significantly improves the constraints on the EDE parameters, while other high redshift data like the Quasar Hubble diagram and the Lyman-$\alpha$ forest BAO have instead a negligible impact. 
We find $\Omega_{\rm eDE} < 0.0039 $ and $w_{0} < -0.95 $ at the $95 \%$ C.L. for EDE accounting for its clustering through the inclusion of perturbation dynamics. This limit becomes weaker $\Omega_{\rm eDE} < 0.0034$ if perturbations are neglected. The constraints on the EDE parameters are remarkably stable even when $\Sigma m_{\nu}$, and $r$ parameters are varied, with weak degeneracies between $\Omega_{\rm eDE}$  and $r$ or $\Sigma m_{\nu}$. In general we expect smaller values for the upper limits on the total amount of EDE with an increasing neutrino mass, while with a decreasing value of the tensor to scalar ratio we expect the 2$\sigma$ upper limits on EDE to increase. We compare this EDE model with a simple $w$CDM with zero dark energy at early times and we find $\sim1-2\%$ different upper limits on total neutrino mass and $\sim0.1-0.2\%$ difference on the equation of state at the present time. 
Perturbation parameters are not constrained with current data sets, and tensions between the CMB derived $H_0$ and $\sigma_8$ values and those measured with local probes are not eased. This work demonstrates the capability of CMB probes to constrain the total amount of EDE well below the percent level.}

\begin{document}
\maketitle
\flushbottom

\section{Introduction}

\label{sec:intro}
The physics behind the evidence for accelerated expansion remains essentially unknown since its first evidence through Type Ia Supernovae  \citep{Riess_1998,Schmidt_1998}. In spite of the considerable progress of cosmological measurements, the origin of the acceleration, which is related to the dominant amounts of the energy budget of the Universe, has become one of the most important problems in cosmology. The well known concordance model of cosmology, the $\Lambda$ Cold Dark Matter ($\Lambda$CDM), describes Dark Energy (DE) as a vacuum-like energy density in the form of a Cosmological Constant (CC), which has an equation of state parameter $w=p/\rho$ (pressure over density) constant in spacetime and equal to $-1$. The $\Lambda$CDM model, including the amplitude and spectral index of density perturbations generated in the early Universe through Inflation, can fit present observational data from Cosmic Microwave Background (CMB), Baryon Acoustic Oscillations (BAO) and Type Ia Supernovae (SNe). For the latest constraints on the $\Lambda$CDM model, which we use in this paper, see the latest results of the Planck satellite \citep{planck2018-VI}. 

Despite these observational achievements, the $\Lambda$CDM model suffers of the energy scale of the CC being unrelated to any of the other known processes of particle physics, which occur at scales tens of orders of magnitude above. For this reason, alternative models of DE have been introduced, including DE paremetrized as a perfect fluid with a time varying equation of state, $w(a)$, where $a$ represents the cosmological scale factor \citep{Chevallier_2000,Linder_2002}.
Among these models, in this work we focus on Early Dark Energy (EDE)
\citep{Wetterich_2004,Doran_2006}, in which the DE is markedly dynamic, and does not need to be negligibly small with respect to the other components in the early Universe, thus easing the fine-tuning problem \citep{Copeland_2006,Peebles_2003,Caldwell_2009}.
EDE models have been considered in the context of generalized theories of Gravity, where the non-minimal Gravitational sector of the Lagrangian constitutes a boost for the DE dynamics in the very early Universe \citep{Perrotta_1999}.  
The EDE deviates from the CC, and the contribution of this non-negligible amount of energy can be described in terms of its initial density parameter $\Omega_{eDE}$. Together with the time variation, the DE has an effect on the growth of density perturbations which can be parametrized by an effective sound speed, the ratio between the  pressure and the density perturbations $c^{2}_{\rm eff} \equiv \delta p / \delta \rho$, and by the anisotropic stress, described as the viscosity sound speed $c^{2}_{vis}$ \citep{Hu1_1998,Hu2_1998,Basse_2012,Putter_2010,Sapone_2009}.
In this paper we relax the assumption of perfect fluid by considering explicitly in the analysis these variables, and we derive constraints based on a combination of the latest data sets.

The structure of the paper is as follows. Section \ref{2nd} describes the EDE model and the behaviour of the perturbations. Section \ref{data} explains the data sets used in this paper. In section \ref{result} we discuss the effects of the perturbations on the CMB data with and without EDE, and in the section \ref{5th} we present the method and the results as well. Finally in Section \ref{conclude} we draw our conclusions.

\section{EDE models}
\label{2nd}

In this Section we define the EDE framework in terms of background evolution and perturbation behavior. 

\subsection{Background evolution}
An approach to construct DE models consists in modifying the energy momentum tensor $T_{\mu \nu}$  in the right hand side of the Einstein equations accounting for a generalized component with negative equation of state. In this way, a Quintessence scalar field $\phi$ with a potential $V(\phi)$ may describe a late time cosmic acceleration \citep{Caldwell_1998,Tsujikawa_2013}. 
 
Unlike the CC scenario, the equation of state of Quintessence models dynamically varies with time. 

EDE represents the class of models in which the DE contribution to the energy density is relevant already in the early Universe, and it can have an impact both on background evolution of geometrical quantities and on structure formation. The notion of EDE has been introduced by Wetterich (2004) \citep{Wetterich_2004} and subsequently studied in several works by considering different possible effective parametrizations of physical properties of the DE.  Here, we concentrate on the general parametrization by Doran and Robbers (2006) \citep{Doran_2006} However, notice that the number of parameters can in principle be reduced as shown in \citep{Pettorino:2013ia}.

In the latter approach, instead of parametrizing $w(a)$, the fractional DE energy density, $\Omega_{\rm DE}(a)$ is written as 
\begin{equation}\label{eq:DE}
    \Omega_{\rm DE}(a) = \dfrac{\Omega^{0}_{\rm DE}-\Omega_{\rm eDE}(1-a^{-3w_{0}})}{\Omega^{0}_{\rm DE}+\Omega^{0}_{\rm m}a^{3w_{0}} }+\Omega_{\rm eDE}(1-a^{-3w_{0}})\,.  
\end{equation}
Here $\Omega^{0}_{\rm DE}$ and $\Omega^{0}_{\rm m}$ are the fractional energy densities of dark energy and matter today, i.e. when the scale factor is normalized to $a=1$; $w_{0}$ is the equation of state of the dark energy fluid today and we also assume a flat Universe. 
Notice that $\Omega_{\rm eDE}$ becomes constant at high redshifts. The evolution of $\Omega_{\rm eDE}$ is connected to the equation of state $w$ by the following relation: 
\begin{equation}
    \Bigg\{3w-\dfrac{a_{\rm eq}}{a+a_{\rm eq}}\Bigg\}\Omega_{\rm DE}(1-\Omega_{\rm DE})=-d\Omega_{\rm DE}/d\ln a\,.
\end{equation}
Therefore the evolution of $w(a)$ reads 
\begin{equation}
    w(a)=-\dfrac{1}{3[1-\Omega_{\rm DE}(a)]}\dfrac{d ln\Omega_{\rm DE}(a)}{d\ln a}+\dfrac{a_{\rm eq}}{3(a+a_{\rm eq})}\,,
\end{equation}
where $a_{\rm eq}$ is the scale factor at matter-radiation equality and also today's equation of state would be written as $w(a=1)=w_{0}$. In order to track the dominant cosmological component, $w(a)$ behaves differently in three different epochs: during the radiation dominated era, one has $w\sim 1/3$, while during the matter domination epoch, $w\sim 0$; finally at present, $w \sim w_{0}$. 

\begin{figure}[!htbp]
    \centering
    \subfloat{{\includegraphics[width=7.22cm]{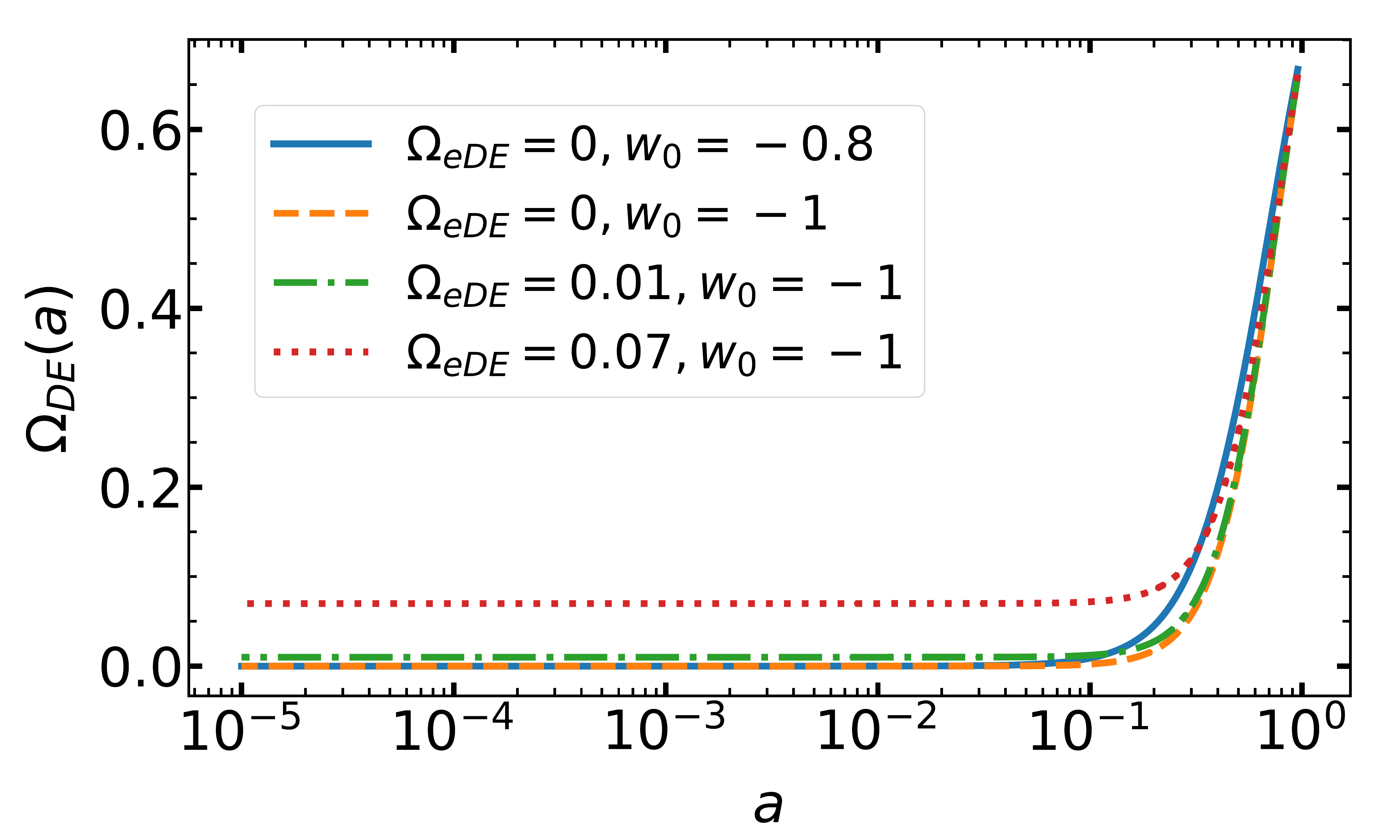} }}
    \qquad
    \subfloat{{\includegraphics[width=7.22cm]{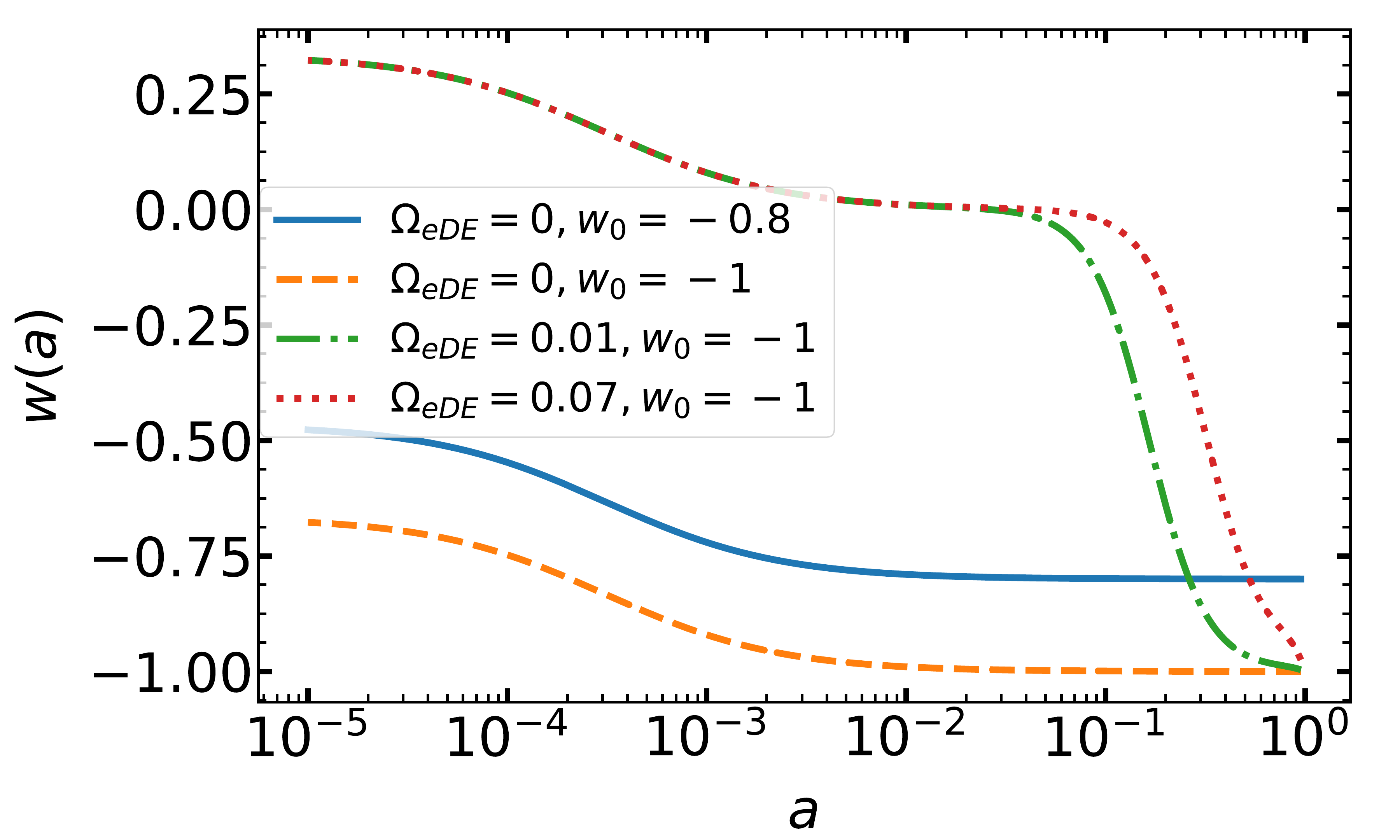} }}
    \caption{The behaviour of the EDE model as a function of scale factor. The left panel shows the evolution of the fractional DE density, while the right panel represents the evolution of the equation of state as a function of the scale factor for different values of $\Omega_{\rm eDE}$ and $w_{0}$.
    }
    \label{fig:example}
\end{figure}
In Figure \ref{fig:example} we plot $\Omega_{\rm DE}(a)$ and $w(a)$, for different values of $\Omega_{\rm eDE}$ and $w_{0}$.

\subsection{EDE perturbations}
\label{3rd}
Besides the background, the additional features that we define and discuss now, make EDE able to influence the behavior of cosmological perturbations. Therefore, DE density perturbations might leave an imprint in cosmological observables. The clustering features of different types of DE models are typically parametrized by an effective sound speed, that can be defined as the ratio of pressure perturbations to the density perturbations in the rest frame of the DE fluid, $c^{2}_{\rm eff} \equiv \delta p / \delta \rho$ \citep{Hu2_1998,Pettorino_2008}. 
In addition, another effective component in the density perturbation of an inhomogeneous DE model is the anisotropic stress which would be considerable for example if the DE behaves like a relativistic fluid with relevant viscosity effects. In order to parametrize the viscosity, we used $c^{2}_{\rm vis}$ as the viscous sound speed \citep{Hu2_1998}.  
As mentioned in Refs.~\citep{Ma_1995,Calabrese_2011,Archidiacono_2014}, by adopting the synchronous gauge in which the perturbation in the metric tensor is confined to the spatial sub-space, and by using the conservation of energy-momentum tensor $T_{\mu \nu}$ in Fourier space, we can have the following relations for density perturbation, velocity perturbation and anisotropic stress:
\begin{equation}\label{pt1}
    \dfrac{\dot{\delta}}{1+w}=-\Bigg[k^{2}+9\Bigg( \dfrac{\dot{a}}{a}\Bigg) \Bigg(c^{2}_{eff}-w+\dfrac{\dot{w}}{3(1+w)(\dot{a}/a)}\Bigg) \Bigg]\dfrac{\theta}{k^{2}}-\dfrac{\dot{h}}{2}-3\dfrac{\dot{a}}{a}(c^{2}_{eff}-w)\dfrac{\delta}{1+w}\,,
\end{equation}

\begin{equation}\label{pt2}
    \dot{\theta}=-\dfrac{\dot{a}}{a}(1-3c^{2}_{eff})\theta+\dfrac{\delta}{1+w}c^{2}_{eff}k^{2}-k^{2}\sigma\,,
\end{equation}

\begin{equation}\label{pt3}
    \dot{\sigma}=-3\dfrac{\dot{a}}{a}\Bigg[1-\dfrac{\dot{w}}{3w(1+w)(\dot{a}/a)}\Bigg]\sigma+\dfrac{8c^{2}_{vis}}{3(1+w)}\Bigg[\theta+\dfrac{\dot{h}}{2}+3\dot{\eta}\Bigg]\,\,;
\end{equation}
$\delta$, $\theta$ and $\sigma$ represent the DE density perturbation, velocity perturbation and anisotropic stress, respectively, while $h$ and $\eta$ are the scalar perturbations of the space-space part in the metric, in the synchronous gauge, and "." denotes the derivative with respect to conformal time. As we will see in the following, the equations above effectively control the DE clustering properties. 

\section{Data sets}
\label{data}

In this section we present the CMB and LSS data sets we exploited in this work. 

\subsection{CMB}\label{planck}

The analysis of Planck data follows dedicated pipelines for the so called low-$\ell$ and high-$\ell$, corresponding to angular scales larger or smaller than a few degrees, respectively. The details of the analyses and data sets are contained in the original publications by Planck \citep{planck2018-V}. We describe here their main features and properties, which are relevant in our context here. 

\subsubsection{Planck-2018 low-$\ell$ data}
For what concerns the low-$\ell$s, following Planck 2018 \citep{planck2018-V}, the baseline low-$\ell$ likelihood adopted in the 2018 legacy release exerts the combination of the following three functions. The first one is a Gibbs-sampling approach in total intensity ($TT$-low-$\ell$ likelihood) and is based on the Bayesian posterior sampling framework which has been implemented by the COMMANDER code \citep{planck2013-XV, planck2015-XI} that has been used extensively in the Planck releases. The second one relies on the estimation of cross-spectra based on the High Frequency Instrument (HFI) channels, 100 and 143 GHz, extended to polarization and including the subtraction of the main diffuse Galactic foreground contamination \citep{planck2016-XLVI}.  
The third function is an updated version of a pixel based likelihood using both total intensity and polarization for $l \leq 29$ \citep{planck2015-XI}; it is based on the 70 GHz Planck channel of the Low Frequency Instrument (LFI), where the diffuse Galactic foregrounds in polarization have been subtracted using the 30 GHz and 353 GHz maps. 

\subsubsection{Planck-2018 high-$\ell$ data}
At high-$\ell$s, the 2019 Planck likelihood corresponds to those used in previous releases \citep{planck2015-XI}, and exploits a power spectrum estimation at multipoles $(30 < \ell < 2500)$, using HFI data. It includes nuisance parameters introduced to control residual systematics, and point source contamination. Planck assumes a Gaussian distribution for the data, written as 
\begin{equation}
    -log\mathcal{L}(\hat{C}|C(\theta))=\dfrac{1}{2}(\hat{C}-C(\theta))^{T} \mathbb{C}^{-1} (\hat{C}-C(\theta))+const.\, ,
\end{equation}
where $\hat{C}$ is the data vector and $C( \theta )$ is the model with (cosmological and nuisance) parameters $\theta$ and $\mathbb{C}$  the covariance matrix. 
The data segments exploited in the Likelihood are in the range $30 < \ell < 1200$ for $100 \times 100$ (100 GHz spectra)
and  $30 < \ell < 2000$ for $143 \times 143$, featuring the best accuracy. Concerning the foreground residual contamination, and associated nuisance parameters, the details of the models and uncertainties are given in Planck 2015 \citep{planck2015-XI}, while the covariance matrix $\mathbb{C}$ is described in Planck 2013 \citep{planck2013-XV}. In comparison with the 2015 releases, the Planck analysis improves the treatment of several systematics and foreground effects. For a comprehensive explanation of cut selections, masks, optical beams and binning of data and also the Galactic and extra-Galactic foregrounds, noise models and calibration, see Planck 2018 \citep{planck2018-V}. 

\subsubsection{Planck-2018 CMB lensing data}\label{pl-lensing}
Gravitational lensing of the CMB can considerably improve the constraints on cosmological parameters. The effect of lensing actually is to remap the CMB fluctuations with an almost Gaussian field representing the lensing angle, with a standard deviation of about 2 arcminutes: the anisotropy in a given direction $\hat{n}$, is re-directed onto the new path represented by $\hat{\textbf n}+ \nabla \phi(\hat{\textbf n})$ where $\phi(\hat{\textbf n}) $ is the CMB lensing potential and $\nabla \phi(\hat{\textbf n})$ denotes lensing deflection angle. The Planck-2018 lensing likelihood \citep{planck2018-III} corresponds to the one used in the previous release  \citep{planck2015-XV}, extended to cover the $8 \leq \ell \leq 400$ multipole interval, which might be important to improve the capabilities of CMB lensing to break geometrical degeneracies in the primary CMB anisotropies. The likelihood is approximated as Gaussian with a fixed covariance estimated from simulations, corresponding to 
\begin{equation}
    -2 log \mathcal{L}_{\phi} = \mathcal{B}^{L}_{i}(\hat{C}^{\phi \phi}_{L}-C^{\phi \phi , th}_{L}) \Big[ \Sigma^{-1} \Big]^{ij} \mathcal{B}^{\prime{L}}_{i} (\hat{C}^{\phi \phi}_{\prime{L}}-C^{\phi \phi , th}_{\prime{L}})\,,
\end{equation}
where $\Sigma$ is the covariance matrix and $\mathcal{B}^{L}_{i}$ are the binning functions, see Planck 2018 \citep{planck2018-III} for details. 

\subsection{LSS}

We consider LSS tracers, relevant for the dynamics of perturbation as well as for background: BAO \citep{Beutler_2011,Ross_2015,Alam_2017} including Lyman-$\alpha$ quasar cross/auto correlations \citep{Bautista_2017,Bourboux_2017,Agathe_2019,Blomqvist_2019}, Type Ia supernova \citep{Betoule_2014}, the recent Hubble diagram for Quasars \citep{Risaliti_2015,Risaliti_2019}, and prior on the present value of the Hubble constant $H_{0}$ which we discuss below.

\subsubsection{Baryon Acoustic Oscillations}\label{bao}
BAO represent the imprints of the oscillations of the photon-baryon plasma in the early Universe and they can be used as a standard ruler in the distribution of the structures today corresponding to the size of the sound horizon at baryon drag
\begin{equation*}
r_s(z_{\rm drag})= \int_{0}^{\eta_{\rm drag}} c_s d\eta = \int_{z_{\rm drag}}^{\infty} \frac{c_s}{H(z)}dz \simeq 150 \, \mathrm{Mpc}\,,
\end{equation*}
where $\eta$ is the conformal time, and $c_s$ the sound speed.
A well known feature of the BAO is represented by a bump in the correlation function of the distribution of the same kind of galaxies and as wiggles in the matter power spectrum which is actually the Fourier transform of the correlation function \citep{Eisenstein_1997,Eisenstein_1998,Eisenstein_1998_2,Eisenstein_2003,Eisenstein_2004,Eisenstein_2005,Eisenstein_2006,Eisenstein_2007,Bassett_2009}.
Measurements of BAO from a galaxy sample constrain the angular diameter distance $D_{A}(z)$ and the expansion rate of the Universe $H(z)$, either separately or in combination through the Alcock-Paczynski test \citep{Alcock_1979}. Indeed, the characteristic scale along the line-of-sight, $s_{\parallel}(z)$, provides a measurement of the Hubble parameter through $H(z) = \dfrac{c \, \Delta z}{s_{\parallel}(z)}$ while the tangential mode, $s_{\perp}$, provides a measurement of angular diameter distance $D_{A}(z) = s_{\perp} (1+z) \Delta \theta$ \citep{Bassett_2009}. On the other hand the Alcock-Paczynski test \citep{Alcock_1979} constrains the product of $D_{A}(z) \times H(z)$, more precisely the volume distance, $D_{v}$, defined as:
\begin{equation}
    D_{v}(z) = \Bigg[(1+z)^{2} D^{2}_{A}(z) \dfrac{cz}{H(z)}\Bigg]^{1/3} \, .
\end{equation}
In this work we use the following BAO data sets: the six-degree-Fields Galaxy survey (6dFGS) at effective $z$ ($z_{\rm eff}$) equal to 0.106 \citep{Beutler_2011}; the Sloan Digital Sky Survey (SDSS) Main Galaxy Sample (SDSS-MGS) at $z_{\rm eff}=0.15$ \citep{Ross_2015}; the complete SDSS-III Baryon Oscillation Spectroscopic Survey cosmological analysis of the Data Release (DR)12 galaxy sample which has been divided into three partially overlapping redshift slices centered at $z_{\rm eff}=0.38$, $0.51$, and $0.61$ \citep{Alam_2017}; the measurement of BAO correlations at $z = 2.33$ with Baryon Oscillation Spectroscopic Survey (BOSS) SDSS DR12 $Ly\alpha$-Forest \citep{Bautista_2017}. Moreover, we also include BAO from the complete SDSS-III $Ly\alpha$-quasar cross-correlation combined with $Ly \alpha$ auto-correlation function at $z = 2.40$ \citep{Bourboux_2017} 
and the BAO measurement at $z = 2.34$ from the recent analyses of correlations (auto-correlation and cross-correlation) of Ly$\alpha$ absorption performed by eBOSS DR14 \citep{Agathe_2019,Blomqvist_2019}. The first two data sets measure $D_{v}/r_{s}$, while the others measure $D_{A}(z_{\rm eff})$, $D_{M}(z_{\rm eff})$ (i.e., the comoving angular diameter distance $D_{M} = (1+z)D_{A}$) and $H(z_{\rm eff})$. In all cases, the BAO measurements are modelled as distance ratios, and therefore they provide no direct measurement of $H_{0}$. However, they provide a link between the expansion rate at low redshifts and the constraints that is placed by CMB data at $z \approx 1100$. Therefore, it is essential to combine CMB with BAO, because the latter can break the degeneracies from CMB measurements and can offer tighter constraints on the background evolution of different dark energy or modified gravity models \citep{planck2015-XIV,Xia_2009}. Finally, notice that BAO measurements are largely unaffected by the non-linear evolution of structures because the acoustic scale is considerably large. 

\subsubsection{Supernovae}\label{sne}
SNe are known to be the most important probe of the accelerated expansion of the universe and DE behavior \citep{Riess_1998,Perlmutter_1999}. They provide accurate measurements of the luminosity distance as a function of $z$. However, the absolute luminosity measurements of SNe is considered to be uncertain and it is marginalized out, removing any constraints on $H_{0}$. Here we used the analysis of the SNe type-Ia by the Joint Light Curve Analysis (JLA) \citep{Betoule_2014}, which is actually constructed from the Supernova Legacy Survey (SNLS)
and SDSS supernova data together with the low redshift supernova data sets.

The motivation for using JLA supernovae dataset rather than the more updated Pantheon dataset \cite{Scolnic:2017caz} is twofold. First, the JLA dataset was used in the Planck 2015 paper \citep{planck2015-XIV} on dark energy and modified gravity. Using JLA allows us to directly compare our constraints on early dark energy parameters (third column of Table \ref{EDEtowcdm} ("EDE fixed $c^{2}_{\rm eff}$ \& $c^{2}_{\rm vis}$") with those from the 2015 Planck paper (last column of Table 3), with the only difference of the updated BAO dataset of our analysis. The second reason is that the quasar dataset that we used in some of our runs is calibrated against the JLA supernova dataset, and, thus, using JLA we are fully consistent. Anyway, we also checked that if we replace the JLA with the Pantheon supernovae dataset the changes in our results are not statistically significant.

\subsubsection{Quasars}\label{qso}
Our analysis includes the Hubble diagram for Quasars (QSOs) as described in Risaliti et al. \citep{Risaliti_2015}, where the constraining power is based on the non-linear relation between the ultraviolet (UV) and X-ray luminosity ($L_{X}$)of QSOs. Where the $L_{X}-L_{UV}$ relation is parametrized as a linear dependence between the logarithm of the monochromatic luminosity at 2500 \AA  ($L_{UV}$) and the $\alpha_{OX}$ parameter defined as the slope of a power law connecting the monochromatic luminosity at $2$ keV ($L_{X}$), and $L_{UV}$: $\alpha_{OX}=0.384 \times \log(L_{X}/L_{UV})$. Luminosities are derived from fluxes through a luminosity distance calculated adopting a standard $\Lambda$CDM model with the best estimates of the cosmological parameters $\Omega_{M}$ and $\Omega_{\Lambda}$. When expressed as a relation $X$-ray and $UV$ luminosities the $\alpha_{OX}-L_{UV}$ relation becomes: 

\begin{equation}
    \log(L_{X}) = \beta + \gamma \log(L_{UV})\,.
\end{equation}   
By using the definition of flux as $F = {L}/(4\pi \, D^{2}_{L})$, the theoretical relation for the $X$-ray is 
\begin{equation}\label{fx}
    \log(F_{X})= \Phi(F_{UV},D_{L}) = \gamma \, \log(F_{UV}) + \Bigg[\bigg[\beta + (\gamma-1)\,\log(4\pi)\bigg]+2(\gamma-1)\,, \log(D_{L})\Bigg]
\end{equation}
where $D_{L}$ is the luminosity distance which, for a $\Lambda CDM$ model with a fixed cosmological constant $\Lambda$, is given by 
\begin{equation}
    D_{L}(z,\Omega_{M},\Omega_{\Lambda})= \dfrac{(1+z)}{\sqrt{\Omega_{K}}} \sinh \sqrt{\Omega_{K}} \times \int_{0}^{z} \dfrac{dz}{H_{0} \sqrt{\Omega_{M}(1+z)^{3}+\Omega_{\Lambda}+\Omega_{K}(1+z)^{2}}} \,,
\end{equation}
With $\Omega_{K}=1-\Omega_{M}-\Omega_{\Lambda}$. By minimizing the likelihood function $(LF)$, one can actually fit the equation \ref{fx} as follows:
\begin{equation}
    \ln(LF) = \Sigma^{N}_{i=1}\Bigg\{ \dfrac{[\log(F_{X})_{i} - \Phi(F_{UV},D_{L})_{i}]}{s^{2}_{i}} + \ln(s^{2}_{i}) \Bigg\} \,, 
\end{equation}
where $s_{i}$ is the error,  $s^{2}_{i}=\sigma^{2}_{2}+\delta^{2}$, with $\sigma_{i}$ and $\delta$ indicating the measurement errors over $F_{X}$ and the global intrinsic dispersion, respectively. We note that the dispersion $\delta$ is much higher than typical values of $\sigma_{i}$. And $N$ is the number of QSOs, here $N=1598$.

\begin{figure}[!htbp]
\centering
\includegraphics[width=120mm]{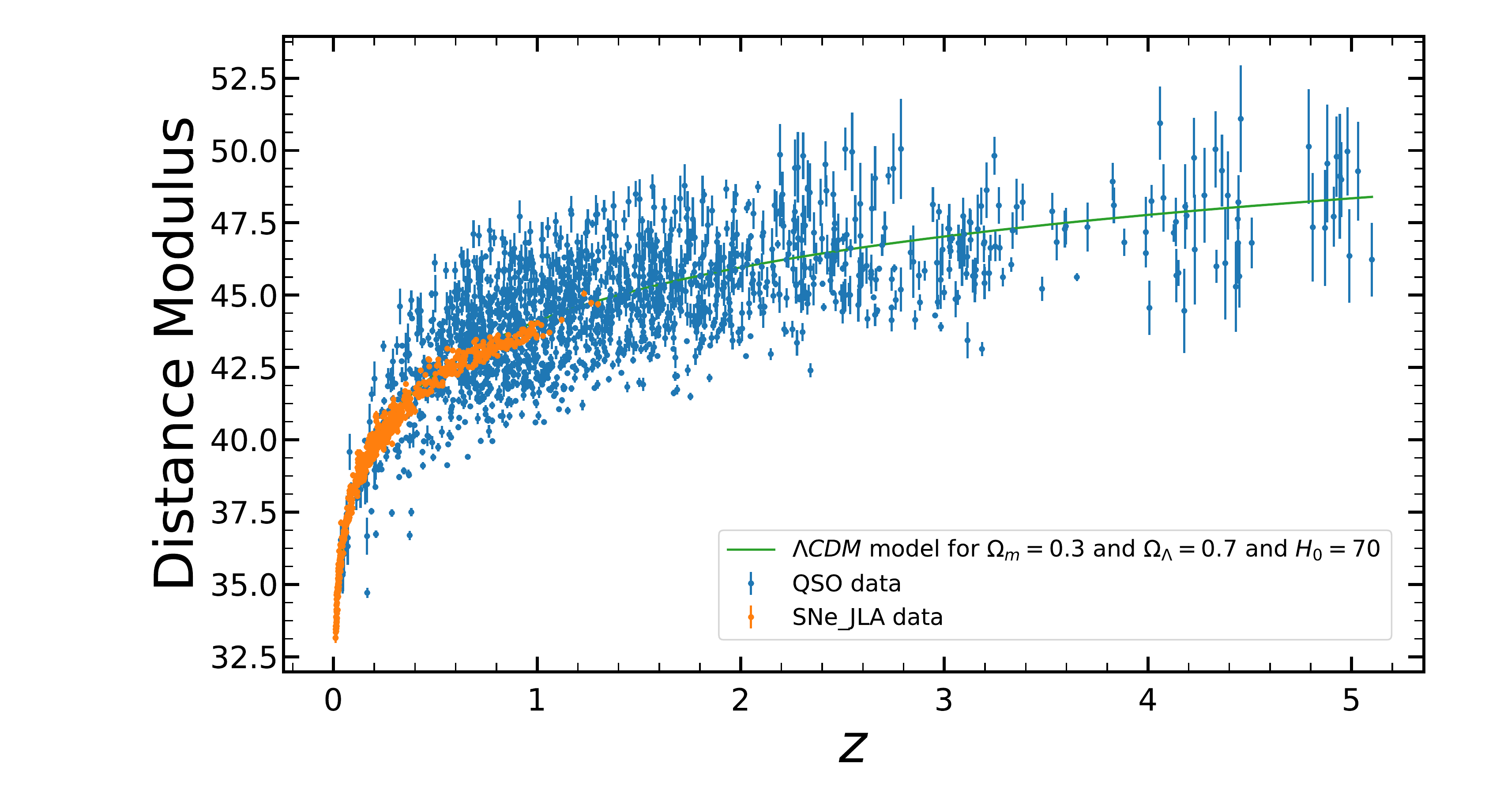}
\caption{The Hubble diagram of supernovae from JLA survey (orange points) and quasars (blue points). The green curve shows the theoretical flat $\Lambda$CDM model with $\Omega_{M}=0.3$, $\Omega_{\Lambda}=0.7$ and  $H_{0}=70$ km/s/Mpc.}
\label{fig2}
\end{figure}

In this paper we used the quasars data points at high redshifts from the recent $1598$ quasar data set in the redshift range corresponding to $0.03<z<5.1$ \citep{Risaliti_2019}. In Figure \ref{fig2} we plot the Hubble diagram of supernovae from the JLA survey and QSO data, showing the markedly different ranges probed and therefore the different constraining power of these two observables.

\subsubsection{The Hubble Constant}\label{h0}
As discussed by Planck 2015 \citep{planck2015-XIV}, dark energy and modified gravity are poorly affected by the physics of recombination, the main influence coming from the integrated Sachs-Wolfe effect and CMB lensing. Following the same reasoning, we use here a re-analysis of the Riess et al. 2011 \citep{Riess_2011} Cepheid data made by Efstathiou et al. 2014  \citep{Efstathiou_2014}. By using a revised geometric maser distance to NGC 258 from Humphreys et al. 2013 \citep{Humphreys_2013}, Efstathiou et al. 2014 \citep{Efstathiou_2014} obtain the following value for the Hubble constant which we adopt (unless specified otherwise) as a conservative $H_{0}$ prior throughout this paper: 
\begin{equation}
    H_{0}=(70.6 \pm 3.3)\, \mathrm{km/s/Mpc} \, .
\end{equation}

Let us point out that, being very broad, this prior is consistent within $1\sigma$ both with the recent direct measurements by Riess et al. 2019 \citep{Riess_2019} and with Planck 2018. The motivation for choosing this particular $H_{0}$ prior is that it is broad enough to be compatible both with Planck and with supernovae data. Moreover, it has been used in Planck 2015 dark energy and modified gravity paper \citep{planck2015-XIV}, and, thus, it is suitable for comparison.

\section{Impact of EDE on cosmological observables}
\label{result}

The goal of this Section is to present the effect that a given amount of EDE has on the main observables considered here: the CMB. We will phenomenologically describe the effect of varying $c^{2}_{\rm eff}$ and $c^{2}_{\rm vis}$ from 0 to 1, separately for the cases in which the EDE component is present or not, i.e. with $\Omega_{\rm eDE}\ne 0$ or $=0$. We will then single out the Integrated Sachs Wolfe (ISW) contribution and the effect on CMB lensing. Finally, we will also show the impact on the linear matter power spectrum, a quantity which is however not used in the present analysis, in order to see the implications that this model could have in terms of the rms value of the amplitude of density fluctuations at 8 Mpc$/h$, corresponding to the $\sigma_{8}$ density parameter. 

The CMB angular power spectrum can be written as the covariance of the total intensity fluctuations in harmonic space:
\begin{equation}\label{eq-cl}
    C_{l} = 4\pi \int \dfrac{dk}{k} \,\mathcal{P_{\chi}} \,|\Delta_{l}(k,\eta_{0})|^{2} \, ,
\end{equation}
where $\mathcal{P_{\chi}}$ is the initial power spectrum and $\eta_{0}$ is today's conformal time. Here $\Delta_{l}(k,\eta_{0})$ is the transfer function for photons, which has the following form on large scales:  
\begin{equation}\label{eq-power}
    \Delta_{l}(k,\eta_{0}) = \Delta^{LSS}_{l}(k) + \Delta^{ISW}_{l}(k) \, ,
\end{equation}
where $\Delta^{LSS}_{l}(k)$ is the contribution of the last scattering surface given by the ordinary Sachs-Wolfe (SW) effect and the total intensity anisotropy and $\Delta^{ISW}_{l}(k)$ is the contribution of the ISW effect. The latter is due to the time change of the potential $\phi$ along the line of sight as follows:
\begin{equation}
    \Delta^{ISW}_{l}(k)=2\int\, d\eta\, e^{-\tau(\eta)} \, \dot{\phi}\, j_{l}[k(\eta-\eta_{0})]\, ,
\end{equation}
where $\tau (\eta)$ is the optical depth coming from the scattering of photons along the line of sight, $j_{l}(x)$ is the spherical Bessel function and $\dot{\phi}$ is the derivative of the potential with respect to the conformal time. 
As already discussed in the literature (see e.g. in Bean $\&$ Dore 2004 \citep{Bean_2004} and Weller $\&$ Lewis 2003 \citep{Weller_2003}),  perturbations in the DE density component with a constant equation of state have a large effect on the largest scales probed by the CMB. 

\subsection{Perturbation effects on the CMB angular power spectrum}

\begin{figure}[!htbp]
    \centering
    \centering
    \subfloat{{\includegraphics[width=7.2cm]{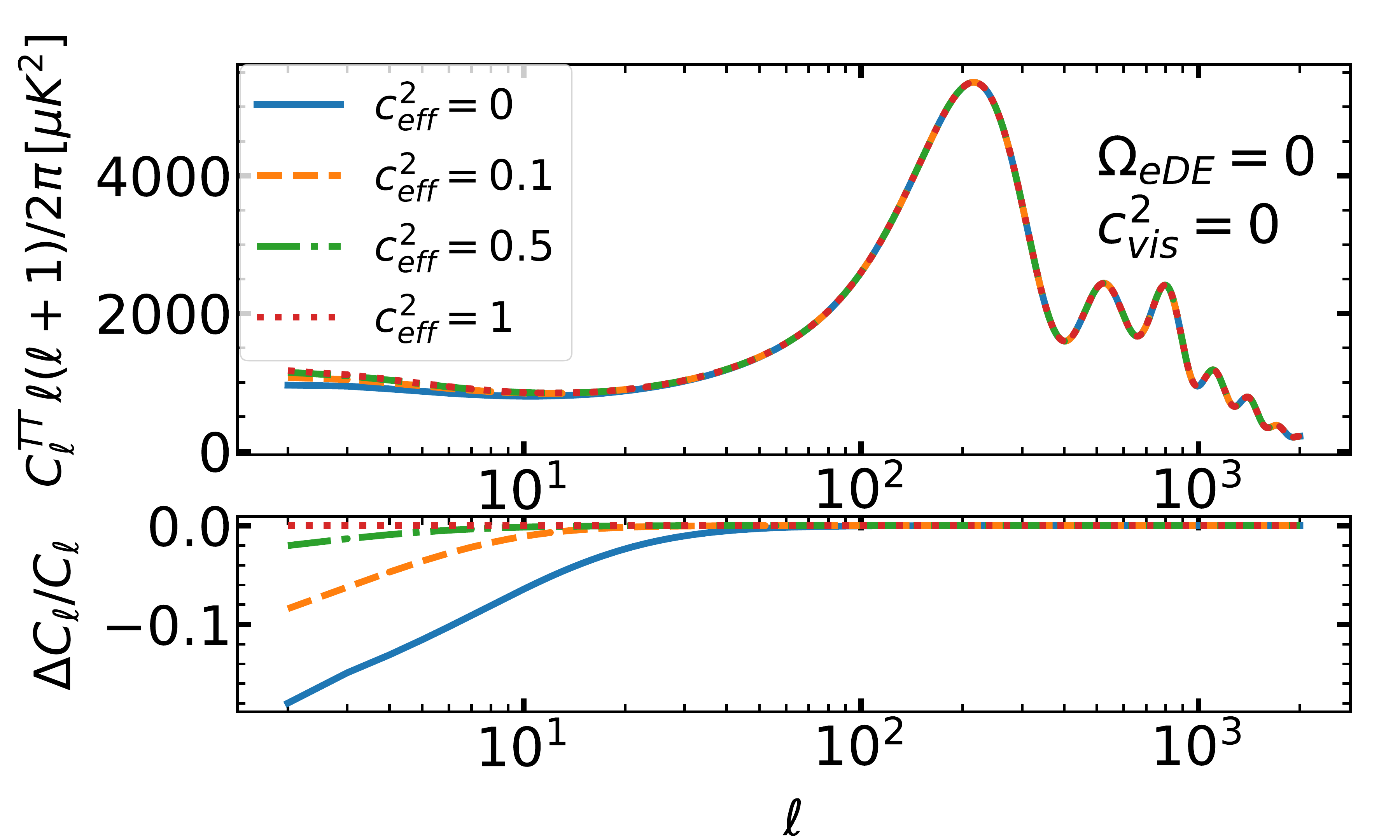} }}
    \qquad
    \subfloat{{\includegraphics[width=7.2cm]{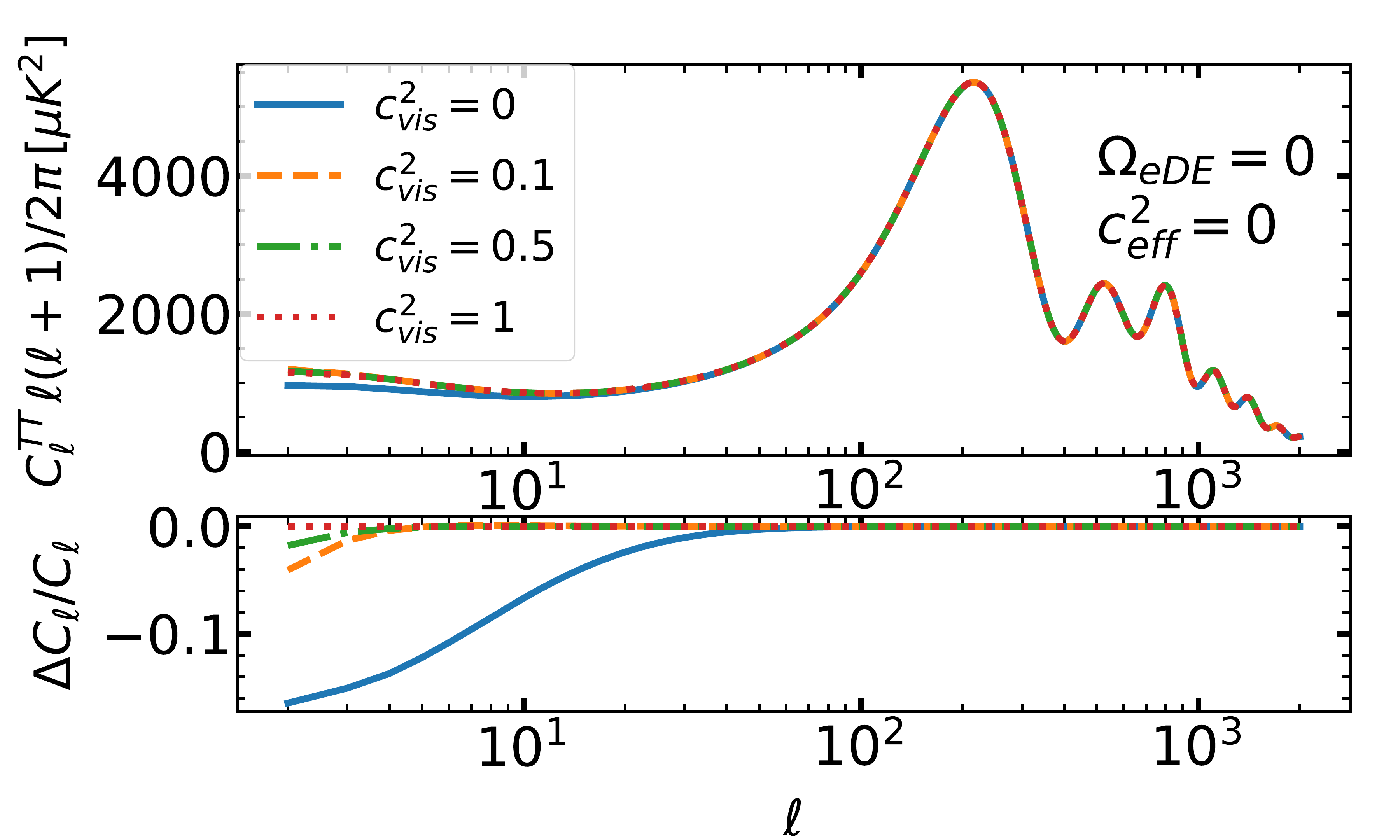} }}
    \caption{The effect of perturbations on the $TT$ CMB angular spectrum for fixed values of $\Omega_{\rm eDE}=0$ and $w_{0}=-0.8$. Note that we use $w_{0}=-0.8$ to make the difference visible, although $w_{0}=-0.8$ is already excluded by data.
    In the left panel the value of the viscous sound speed has been fixed, $c^{2}_{\rm vis}=0$, in order to see the effect of varying the effective sound speed $c^{2}_{\rm eff}$ on the CMB angular power spectra. In the right panel the value of the effective sound speed has been fixed, $c^{2}_{\rm eff}=0$, to see the effect of varying the viscous sound speed $c^{2}_{\rm vis}$. The relative effects of each case has been shown in the lower panels.
    }
    \label{cmb1}
\end{figure}

Following the perturbation equations \ref{pt1}, \ref{pt2} and \ref{pt3}, we plot the CMB angular power spectrum for different values of $c^{2}_{\rm eff}$ and $c^{2}_{\rm vis}$ in Figure \ref{cmb1}. Note that here we are switching off the presence of EDE, i.e. $\Omega_{\rm eDE}=0$, in order to investigate phenomenologically  the pure effect of perturbations.

\begin{figure}[!htbp]
    \centering
    \subfloat{{\includegraphics[width=7.2cm]{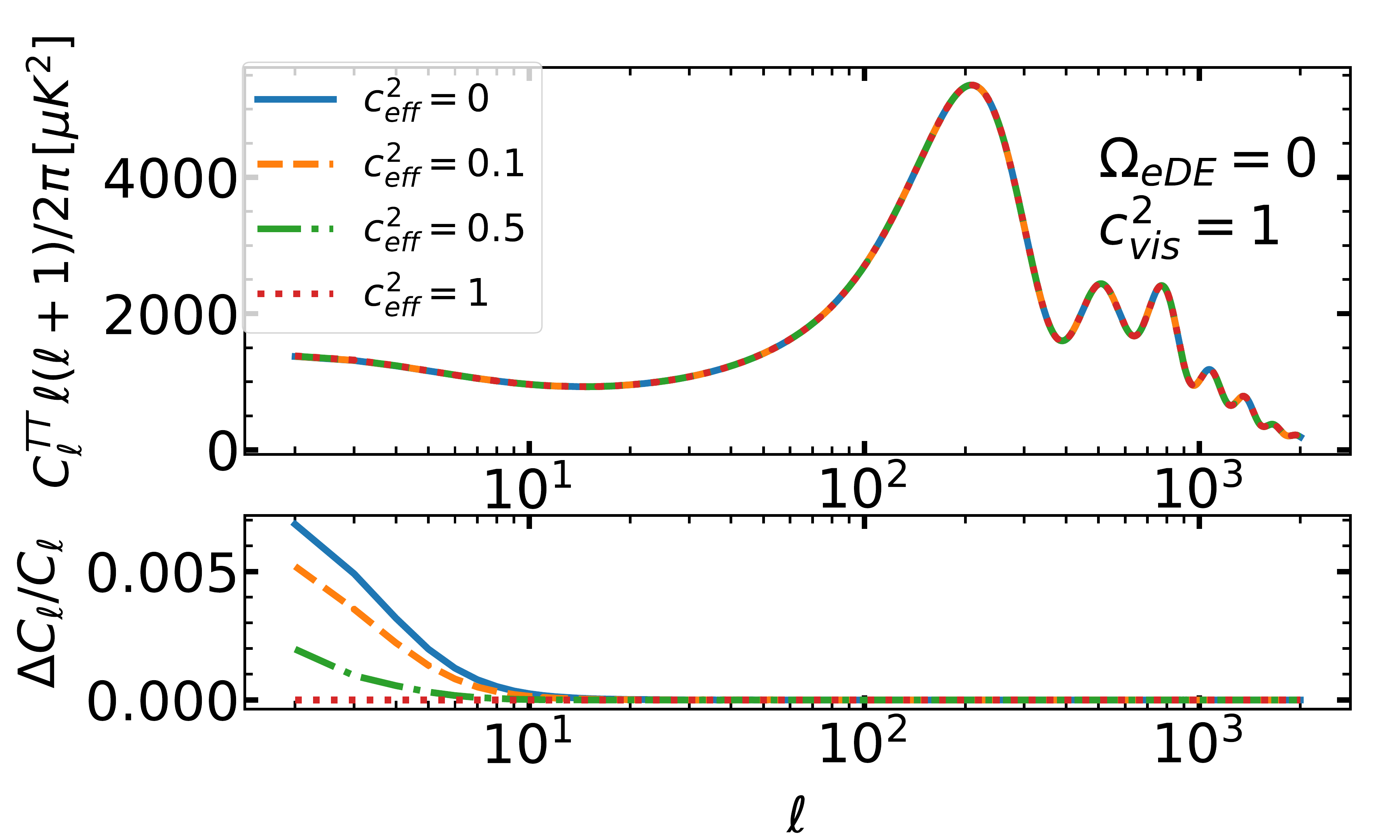} }}
    \qquad
    \subfloat{{\includegraphics[width=7.2cm]{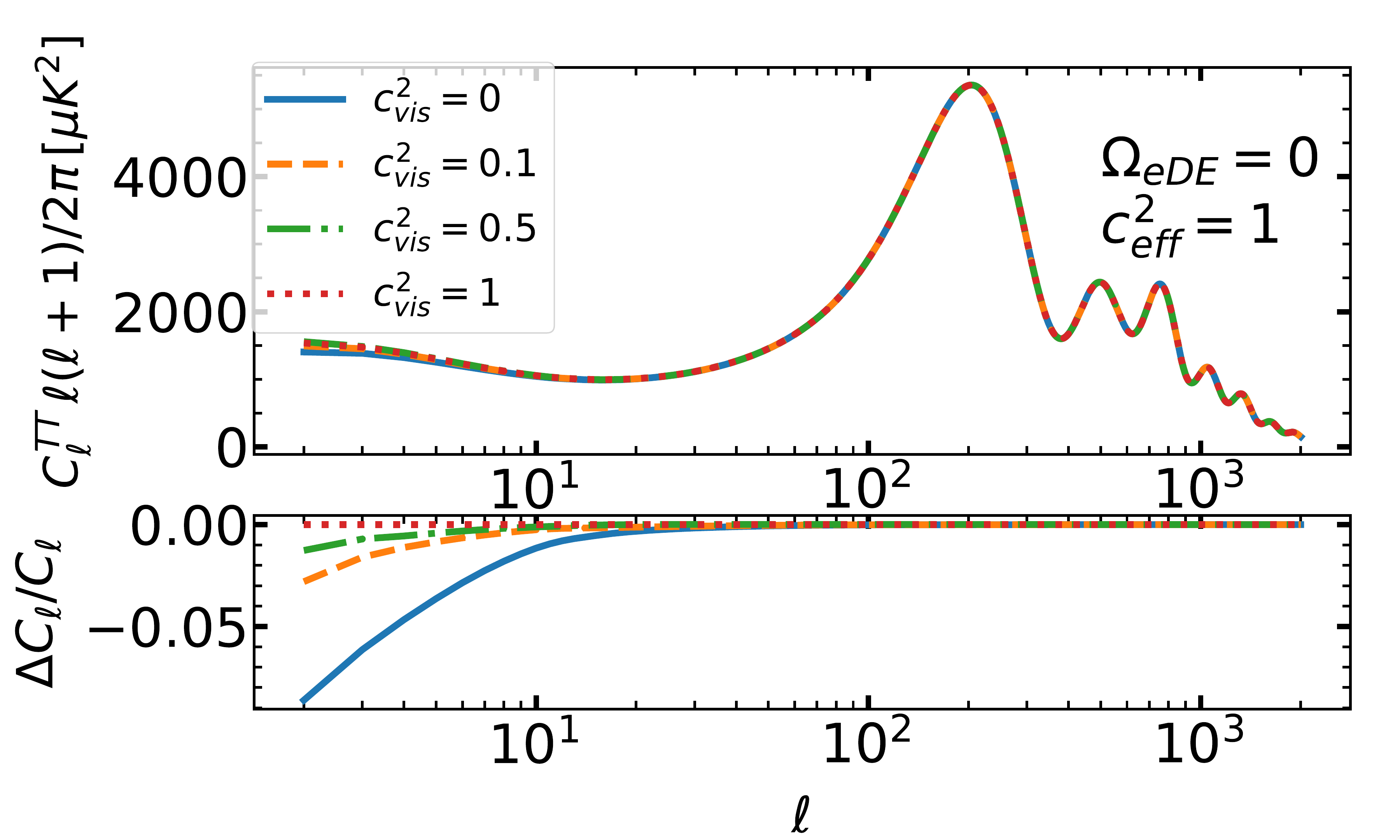} }}
    \caption{The effect of perturbations on the $TT$ CMB angular power spectrum for the value of $\Omega_{\rm eDE}=0$. The value of the viscous sound speed has been fixed, $c^{2}_{\rm vis}=1$ to see the effect of the effective sound speed $c^{2}_{\rm eff}$ on the CMB angular power spectra (left panel). The value of the effective sound speed has been fixed, $c^{2}_{\rm eff}=1$ to see the effect of the viscous sound speed $c^{2}_{\rm vis}$ in the right panel.
    }
    \label{cmb2}
\end{figure}

As it can be seen in Figure \ref{cmb1}, the effect on the CMB spectrum are only confined at relatively large scales due to the ISW effect. The reason would be the fact that since in the scenario with the constant equation of state and a negligible energy component in the early Universe which means: ($w(a)=w_{0}$ and $\Omega_{eDE}=0$), the dark energy has a contribution in the energy density only at late times. Therefore the CMB power spectrum can be only influenced by the late ISW effect. The effect that would be achieved by increasing $c^{2}_{\rm eff}$ or $c^{2}_{\rm vis}$ is the higher ISW power. And this fact replies the increased potential caused by the dark energy. Although the dark energy perturbation would help to keep the potential constant, increasing $c^{2}_{\rm eff}$ or $c^{2}_{\rm vis}$ can reduces the dark energy perturbation and this fact leads to diminish the decay of the potential. And by the decay of potential the ISW power increases. In the left panel of the Figure \ref{cmb1}, by fixing $c^{2}_{\rm vis}=0$ and increasing the value of $c^{2}_{\rm eff}$ gradually from $0$ to $1$, as discussed above, the dark energy perturbation contribution decreases and leads to the decay in potential, Therefore the ISW effect increases and due to the equation \ref{eq-power} the transfer function of photons increases as well, so as can be seen in the equation \ref{eq-cl}, the CMB angular power spectrum increases. 
In comparison, the same effect is happening in the right panel, by fixing $c^{2}_{\rm eff}=0$ and increasing the amount of $c^{2}_{\rm vis}$ gradually from $0$ to $1$. In this way the perturbations are suppressed and the ISW effect increases as a consequence of the dynamics deriving from the suppression itself.
Therefore, in both cases, by fixing one parameter and increasing the value of the other,  we have an increase in the amount of ISW component and the CMB angular power spectrum as well. As already discussed in the literature, the feasibility of accurately measuring one of these parameters is strongly undermined by the presence of cosmic variance on the angular scales in which the ISW is effective. \\
We now fix one of the parameters to $1$ and increase the other parameter gradually from $0$ to $1$. As expected, and shown in Figures \ref{cmb1} and \ref{cmb2}, by fixing one perturbation parameter and changing the other we observe an effect which is similar to the one of perturbations on the CMB angular power spectra: the effect of increasing $c^{2}_{\rm eff}$ or $c^{2}_{\rm vis}$ reduces the DE perturbations and this can leads to the decaying of the potential and therefore to a larger ISW effect. For all the cases, the impact is mostly seen at large scales, at multipoles $l< 80$ and bound to be below the 3-4\% level. In the following Section we are going to check the effect of the early dark energy model on the CMB.

\subsection{Effects of the early dark energy on the CMB angular power spectrum}

It is important now to focus on the effect of a given amount of early dark energy on the CMB. Therefore, in the following figures the combined effect of perturbations and a non-zero energy density in the dark energy fluid can be investigated.

\begin{figure}[!htbp]
    \centering
    \subfloat{{\includegraphics[width=7.2cm]{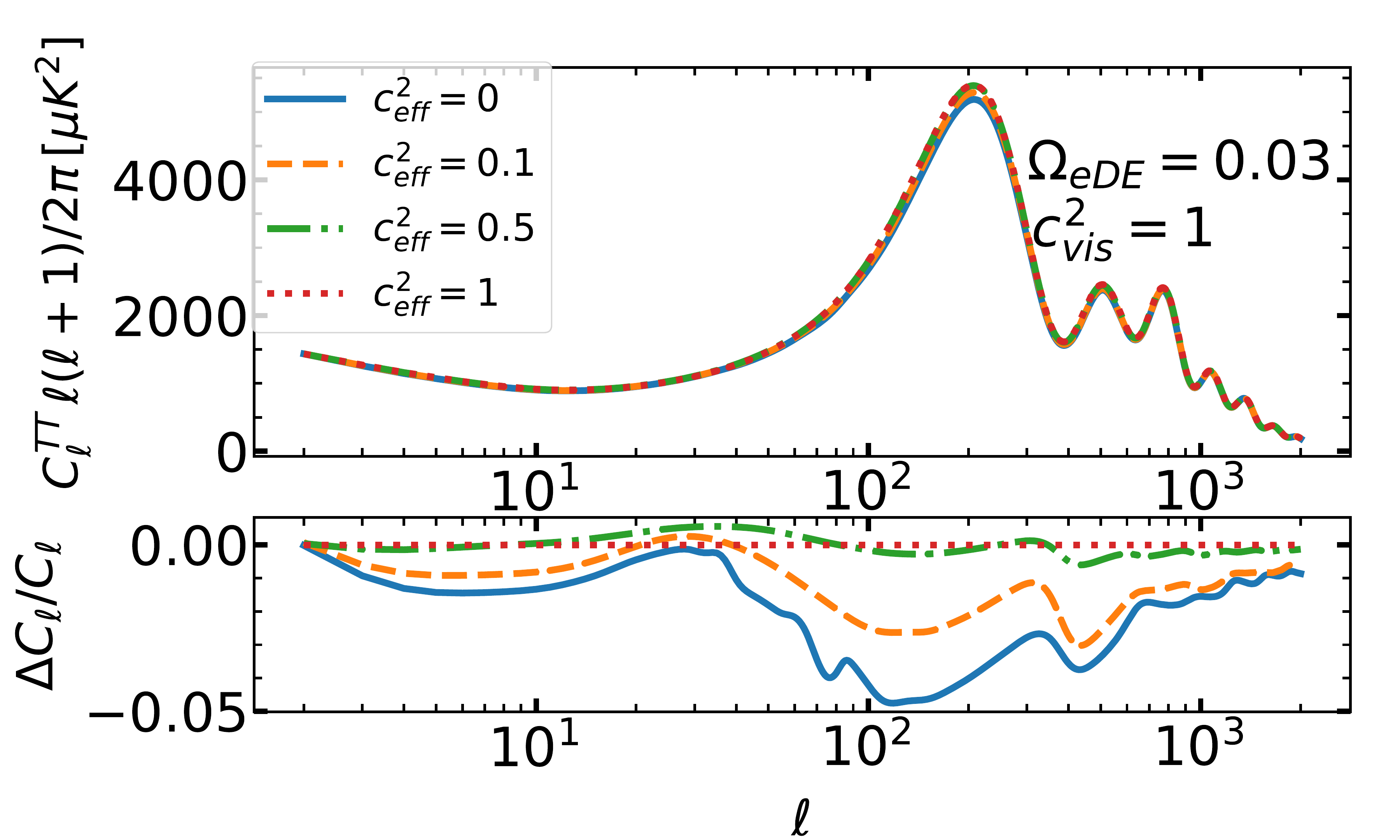} }}
    \qquad
    \subfloat{{\includegraphics[width=7.2cm]{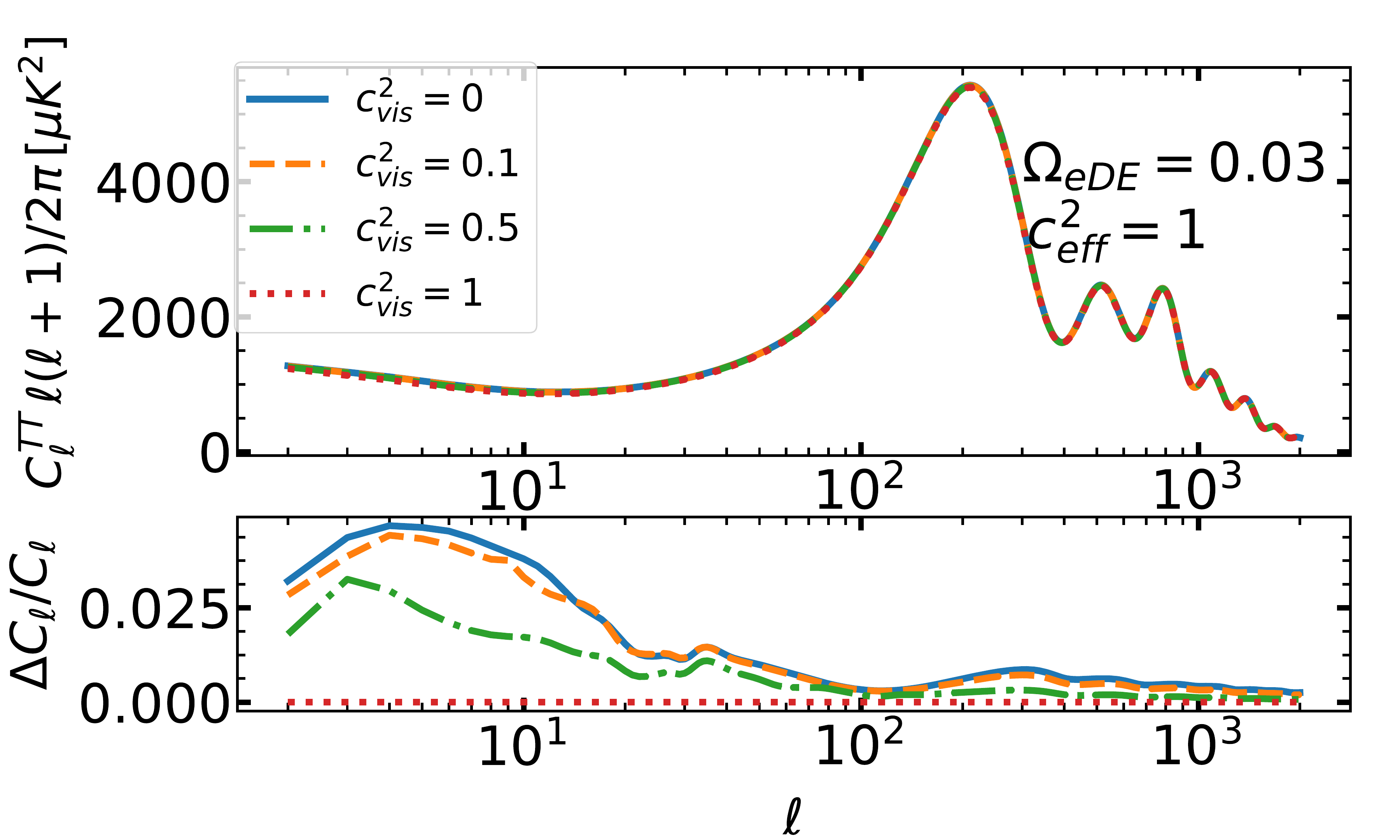} }}
    \caption{EDE effects on the $TT$ CMB angular power spectrum for the indicated value of early dark energy $\Omega_{\rm eDE}=0.03$. Note that, although the value of $\Omega_{eDE}$ chosen here is already excluded by data, we use it to make the difference more visible. The value of the viscous sound speed has been fixed, $c^{2}_{\rm vis}=1$ in the left panel and $c^{2}_{\rm eff}=1$ in the right panel, respectively.
    }
    \label{cmb4}
\end{figure}

\begin{figure}[!htbp]
    \centering
    \subfloat{{\includegraphics[width=7cm]{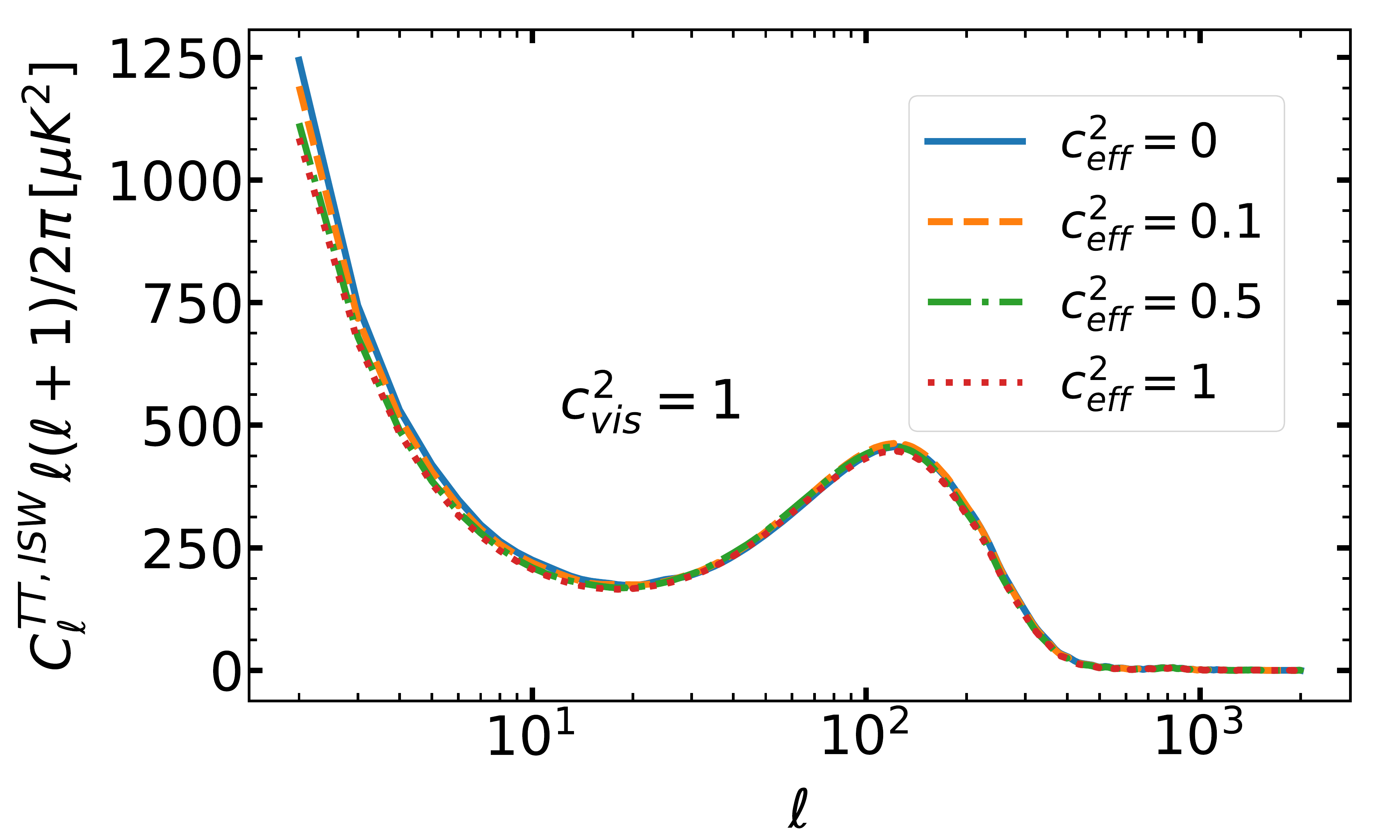} }}
    \qquad
    \subfloat{{\includegraphics[width=7cm]{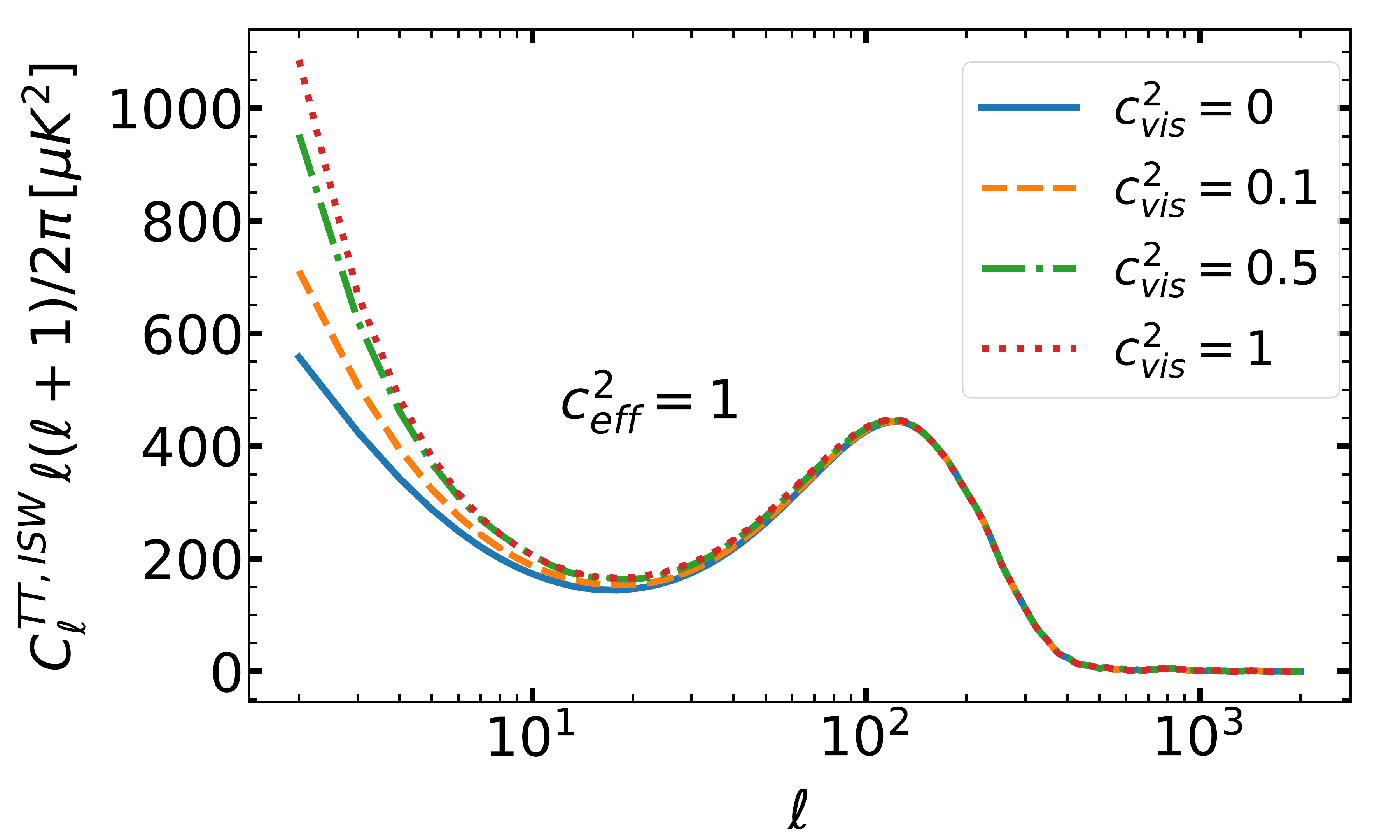} }}
    \caption{The behaviour of the ISW component of the TT CMB angular power spectrum when the EDE effect is on. The values of $\Omega_{\rm eDE}=0.03$ and $w_{0}=-0.8$ are fixed in all the curves. As in previous figures, notice that the values of $\Omega_{\rm eDE}$ and $w_{0}$ are chosen to make the differences more visible, although, as shown in Table \ref{alldata}, these values are already excluded by data. In the left(right) panel the value of the viscose(effective) sound speed has been fixed to see the effect of the effective(viscose) sound speed. 
    }
    \label{isw}
\end{figure}

Figure \ref{cmb4} shows the effects of perturbations when $\Omega_{\rm eDE}\ne 0$. They are visible also on smaller scales with respect to a pure ISW, due to the contribution of the early ISW, associated with a non-zero EDE. The difference is particularly visible for the first acoustic peak. The reason is that, for $\Omega_{\rm eDE}\ne 0$, the EDE influences directly the recombination process so the EDE can affect on the evolution of the acoustic oscillations before recombination. Although the differences are small, but more significant with respect to the ISW, due to the reduced cosmic variance. Finally, we notice that, similarly to the previous case, by increasing the sound speed, perturbations in the EDE get more and more suppressed, leading to a stronger decay of the metric perturbations.
The behaviour of the ISW effect is shown in detail in Figure \ref{isw} which displays only the ISW component of the CMB spectra, highlighting the late ($(l<30)$) and early ($l \sim 120$) contributions. 

\subsection{CMB Lensing}\label{cmb_lensing}
\begin{figure}[!htbp]
\centering
\includegraphics[width=160mm]{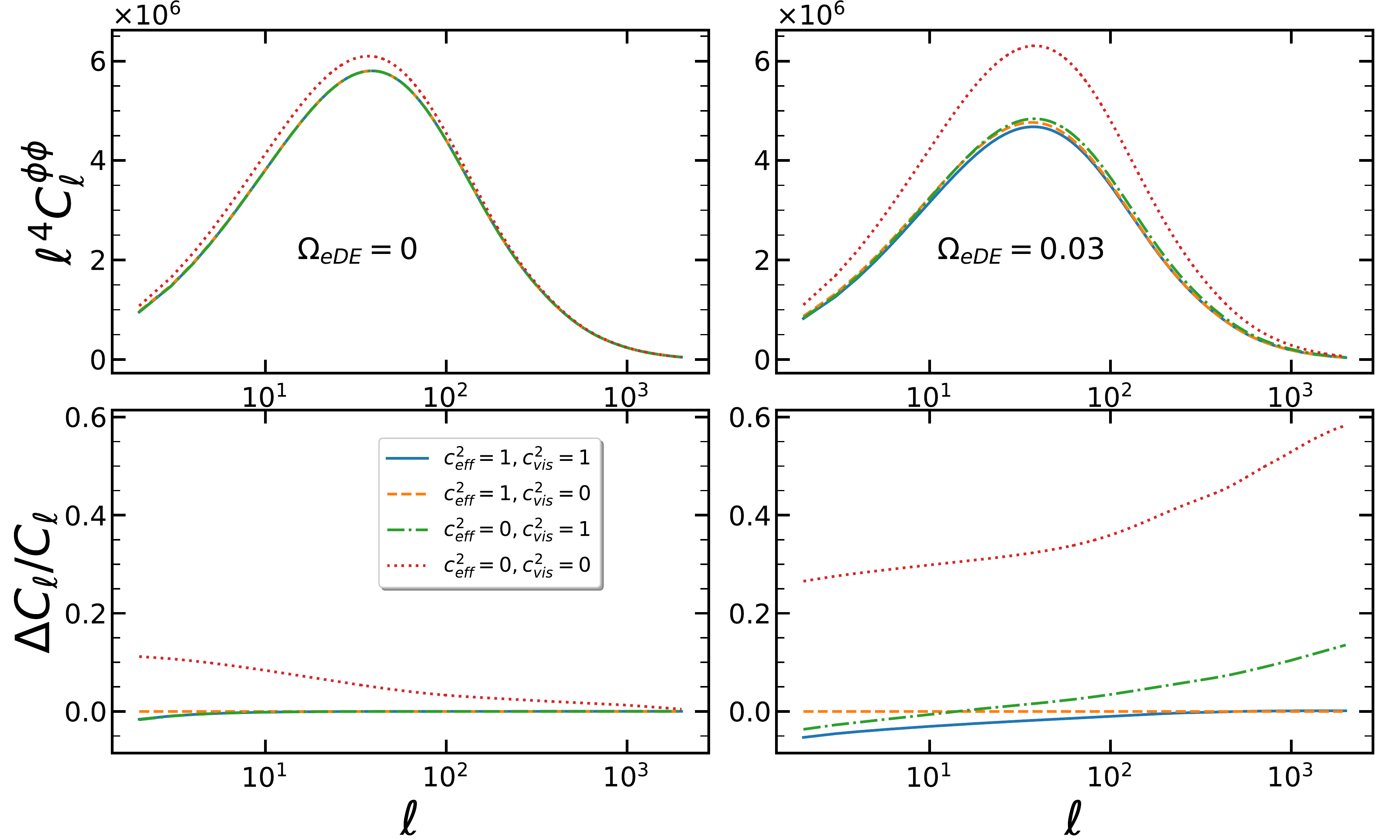}
\caption{Lensing potential power spectrum in scenarios with vanishing or finite $\Omega_{eDE}$ for different values of perturbation parameters ($c^{2}_{\rm eff}$ and $c^{2}_{\rm vis}$). Left panels show the scenarios without early dark energy $\Omega_{eDE}=0$ and the right panels show the lensing potential with the early dark energy parameter $\Omega_{eDE}=0.03$. The ratios with respect to a model with $c^{2}_{\rm vis}=0$ and $c^{2}_{\rm eff}=1$ has been shown in the lower panels; $w_{0}=-0.8$ is fixed and $\Omega_{\rm eDE}=0.03$ when different from zero. We have to specify that the values chosen for $w_{0}$ and $\Omega_{\rm eDE}$ are only meant to make the difference more visible. Table \ref{alldata} will show that these values are actually excluded by data.}

\label{lensing}
\end{figure}

Similarly to the previous Section, Figure \ref{lensing} shows the lensing potential angular power spectra with and without EDE, for different values of the perturbation parameters. In the case $\Omega_{\rm eDE}=0$, one can clearly see that if by taking $c^{2}_{\rm eff}=1$ or $c^{2}_{\rm vis}=1$, perturbations are suppressed, making the lensing potential nearly equivalent in the two cases. In the case with $c^{2}_{\rm eff}=0$ and $c^{2}_{\rm vis}=0$, where no friction is caused to perturbation growth, the lensing potential is significantly enhanced (see the red dotted curve in Figure \ref{lensing}). Similarly, when $\Omega_{\rm eDE}=0.03$, $c^{2}_{\rm eff}$ or $c^{2}_{\rm vis}$ is equal to $1$ causes a suppression onto perturbations, while for $c^{2}_{\rm eff}$ and $c^{2}_{\rm vis}$ equal to $0$, the lensing potential would be significantly enhanced. Thus, $\Omega_{\rm eDE}\ne 0$ causes a stronger enhancement in comparison with  non-EDE scenarios because of the fact that the presence of the EDE leads to a larger DE clustering, causing a more pronounced lensing power. 

\begin{figure}[!htbp]
\centering
\subfloat{\includegraphics[width=72mm]{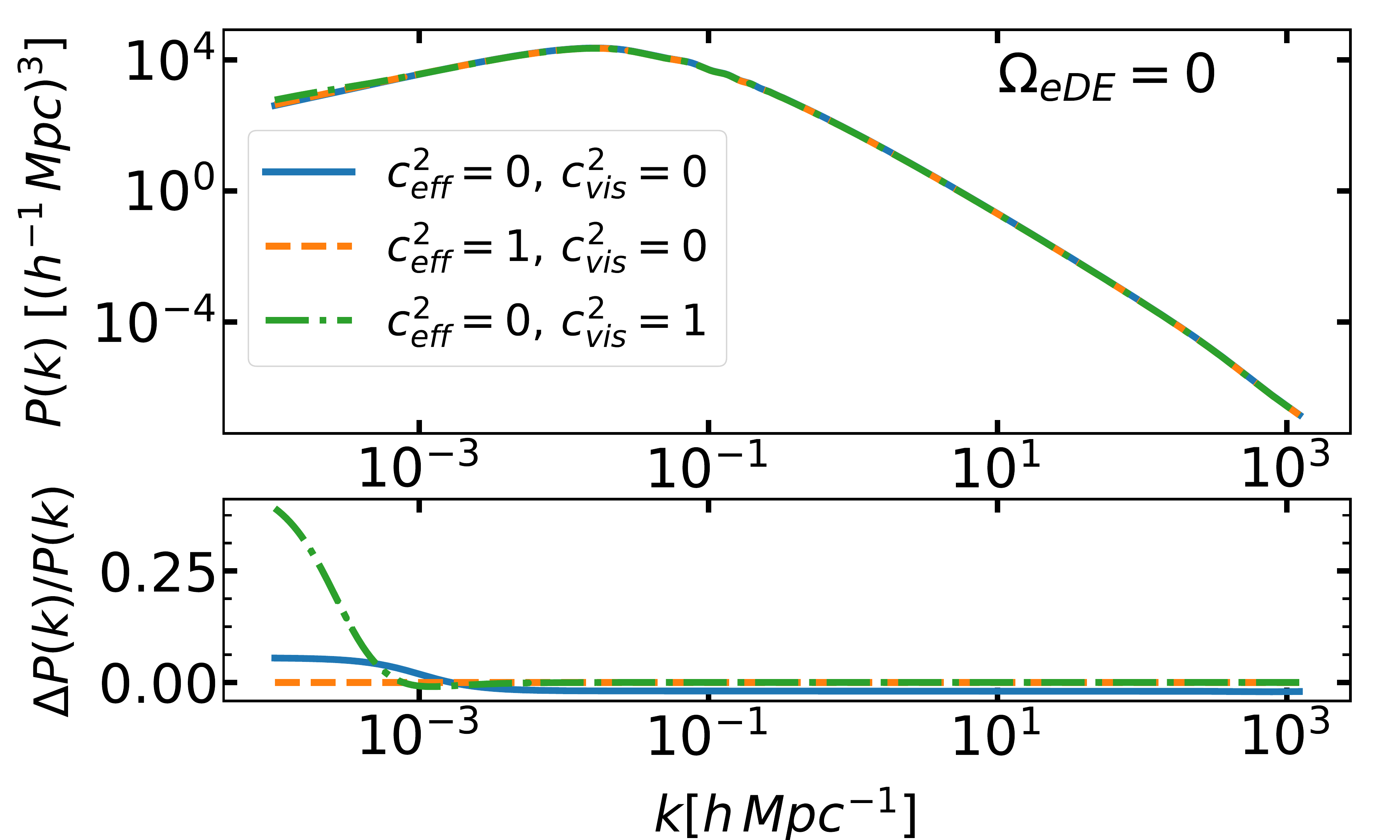}}
\qquad
\subfloat{\includegraphics[width=72mm]{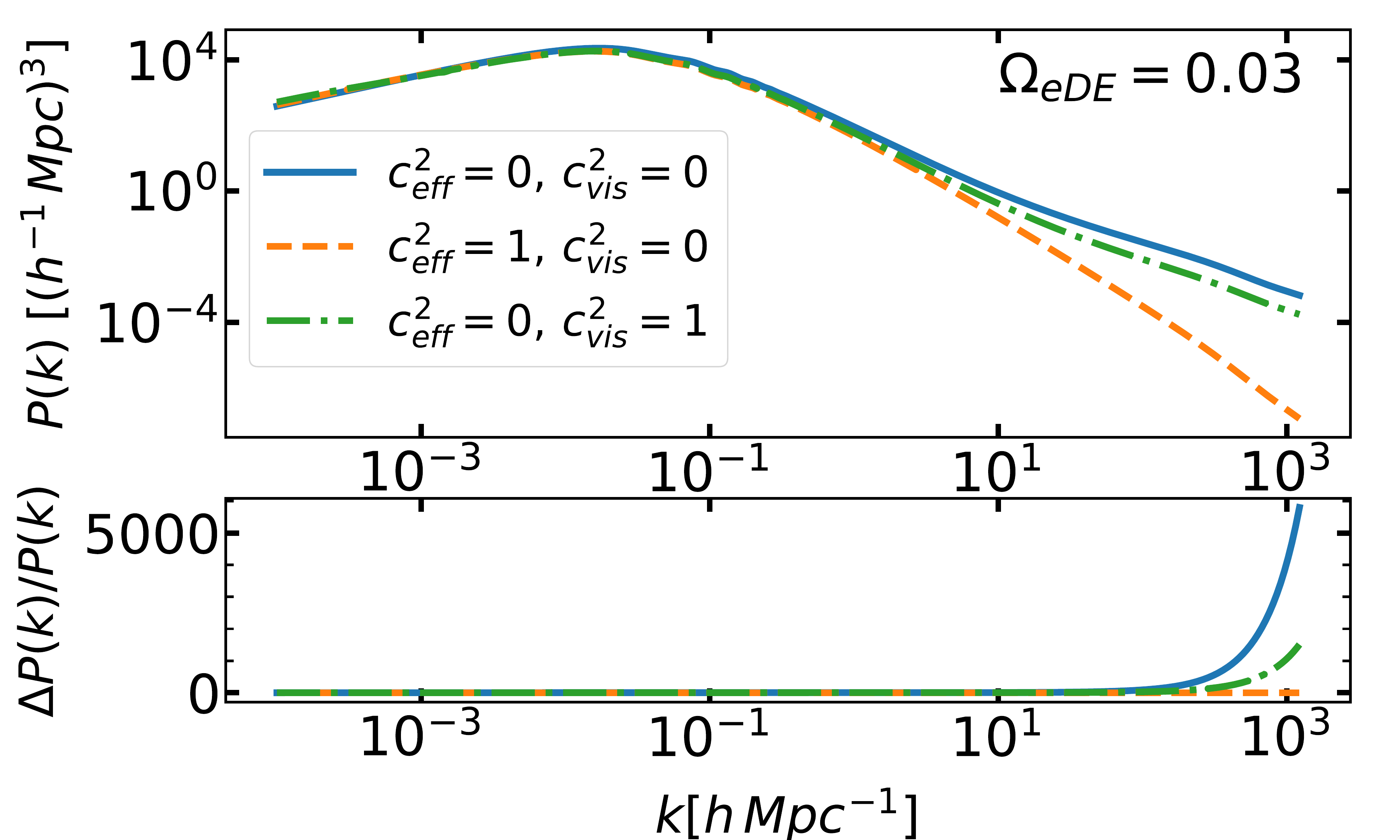}}
\caption{Linear matter power spectrum for $\Omega_{eDE}=0.03$ (right panel) and $\Omega_{eDE}=0$ (left panel) models for $w_{0}=-0.8$ and different values of the perturbation parameters, as indicated. As already mentioned in the captions of previous figures, for the sake of visibility we chose $w_{0}=-0.8$ and $\Omega_{eDE}=0.03$, although these values are already excluded by data.}

\label{power}
\end{figure}

\subsection{Effects on the matter power spectrum}
Figure \ref{power} shows the impact of EDE for different values of the perturbation parameters. The left panel is without the EDE effect and the right panel we have EDE by $\Omega_{\rm eDE}=0.03$. 
The blue solid curve in the right panel, shows the matter power spectrum for $c^{2}_{\rm eff}=0$ and $c^{2}_{\rm vis}=0$, i.e. when perturbations are not affected by friction at all. We can clearly see that there would be a significant enhancement at small scales in comparison with the other combination of the perturbation parameters ($c^{2}_{\rm eff}$ and $c^{2}_{\rm vis}$). Basically this enhancement can be decreases if we take $c^{2}_{\rm eff}=1$ or $c^{2}_{\rm vis}=1$ because of the suppression of the perturbations. In each panel the relative differences form the reference case i.e. ($c^{2}_{\rm eff}=1$ and $c^{2}_{\rm vis}=0$) is plotted as well.

Overall, the impact of the amount of EDE and varying $c^{2}_{\rm eff}$ and $c^{2}_{\rm vis}$ on cosmological observables, mainly CMB here, can be summarize as follows. In the absence of the EDE component ($\Omega_{eDE}=0$), we can see two different effects on the CMB, mainly on the ISW power (see Figures \ref{cmb1} and \ref{cmb2}). If we switch off one of the $c^{2}_{\rm eff}$ or $c^{2}_{\rm vis}$ parameters, by increasing the other parameter we could achieve a higher ISW power about $\sim 10\%$. (Figures \ref{cmb1}). Contrary, if we fix one of the parameters to $1$ and increase the other parameter, the impact depends on which parameter we are fixing: if we fix $c^{2}_{\rm vis}=1$, by increasing the other parameter we would have a decrease in the ISW power by about $\sim 0.5\% $; instead, if we fix $c^{2}_{\rm eff}=1$ and increase the other parameter, we would have $\sim 5\%$ decrease in ISW power (Figure \ref{cmb2}). Therefore it seems that even in the case where there is no EDE component ($\Omega_{eDE}=0$), each parameter has its own effect on the CMB power spectra. So we need both parameters  $c^{2}_{\rm eff}$ and $c^{2}_{\rm vis}$ to describe the ISW effect. By switching on the EDE component, this differences can be visible also on smaller scales (see Figure \ref{cmb4}). \\
Besides primary CMB, we also investigated the impact of switching on and off $c^{2}_{\rm eff}$ or $c^{2}_{\rm vis}$ parameters on the CMB lensing spectra in the presence or in the absence of a non-zero EDE component separately (Figure \ref{lensing}). When $\Omega_{eDE}=0$, fixing either $c^{2}_{\rm eff}=1$ or $c^{2}_{\rm vis}=1$, and setting to 0 the other perturbation parameter does not lead to any significant variation of the lensing spectra. Only if both $c^{2}_{\rm eff}=0$ and $c^{2}_{\rm vis}=0$, then there is no friction in the growth of perturbations, and the lensing potential is significantly enhanced (Figure \ref{lensing}-right panel). On the other hand, in the presence of a non-zero EDE component, even setting to 0 only one of the perturbation parameters leads to a noticeable increase of the lensing power. Still we can see the significant enhancement by switching off both parameters (Figure \ref{lensing}-right panels). \\
We summarized all the effects on the CMB spectrum in Table \ref{my-label}. \\
Finally, we study the impact on the linear matter power spectrum. As can be seen in Figure \ref{power}, each components can have different effects on the linear matter power spectrum both in the absence and in the presence of a non-zero EDE component.


\begin{table}[]
\centering
\caption{The impact of EDE and perturbation parameters on CMB spectrum.}
\label{my-label}
\begin{tabular}{|l|l|l|l|l|}
\hline
\hline
Observables &$\Omega_{eDE}$ & $c^{2}_{\rm vis}$ & $c^{2}_{\rm eff}$ & Effect \\ \hline\hline
ISW        & $=0$ & $0$ & Increasing & $\sim 10\%$ Increase at large scales\\  \cline{3-4} \cline{4-5}  
           &  & Increasing & $0$ & $\sim 10\%$ Increase at large scales\\ \cline{3-4} \cline{4-5}
           &  & $1$ & Increasing & $\sim 0.5\%$ Decrease at large scales\\
           \cline{3-4} \cline{4-5}
           &  & Increasing & $1$ & $\sim 5\%$ Increase at large scales\\
           \cline{2-3} \cline{3-4} \cline{4-5}
           & $\neq 0$ & $1$ & Increasing & $\sim$ up to $4\%$ Increase at all scales\\ \cline{3-4} \cline{4-5}
           & & Increasing & $1$ & $\sim$ up to $4\%$ Decrease at all scales\\  \cline{1-2} \cline{2-3} \cline{3-4} \cline{4-5}
           
CMB lensing        & $=0$ & $0$ & $0$ & $\sim 10\%$ Increase\\  \cline{3-4} \cline{4-5}
                & $\neq 0$ & $0$ & $0$ & $\sim 45\%$ Increase \\  \cline{1-2}\cline{2-3}\cline{3-4} \cline{4-5}
\hline
\end{tabular}
\end{table}

\section{Constraints on spacetime dynamics for EDE}\label{5th}

In this Section we derive the constraints on the EDE scenarios, considering the latest data sets, and the phenomenology outlined above, in a wider parameter space that includes variation in the total neutrino mass $\Sigma m_{\nu}$, and the tensor-to-scalar-ratio $r$.  

\subsection{Methodology and parametrization}

We analyze here the EDE models by using a modified version of the Boltzmann equation solver CAMB \citep{Lewis_2000} in order to account for equations (\ref{eq:DE}) and (\ref{pt1})-(\ref{pt3}) through varying the following set of parameters:
\begin{equation}
    \{ \Omega_{\rm b}h^{2}, \Omega_{\rm c}h^{2}, 100 \theta_{\rm MC}, \ln[10^{10}A_{s}], n_{s}, \tau, \Sigma m_{\nu},  r, \Omega_{\rm eDE}, w_{0}, c^{2}_{\rm eff}, c^{2}_{\rm vis} \} \, .
\end{equation}
We consider the standard six parameters of the concordance $\Lambda$CDM model \citep{planck2018-III}, i.e. the baryon and CDM fractional densities today $\Omega_{\rm b}h^{2}$, $\Omega_{\rm c}h^{2}$, $100$ times of the ratio between the sound horizon and the angular diameter distance at decoupling ($100 \times r_{*}/D_{A}$) which is usually denoted by $100 \, \theta_{MC}$, the primordial scalar perturbations amplitude $\ln[10^{10}A_{s}]$, the scalar spectrum power-law index $n_{\rm s}$, and the reionization optical depth $\tau$. In addition, we include the neutrino masses $\Sigma m_{\nu}$, and the ratio of the tensor primordial power to the scalar curvature one at $k_{0}= 0.05$ Mpc$^{-1}$ which is called $r$. The last four parameters are related to the EDE scenario, in which the possibility of clustering also has been included. As already discussed in Section \ref{2nd}, the parameters could be described as follows: $\Omega_{\rm eDE}$, the non-negligible fractional DE density in the early universe, $w_{0}$, the equation of state parameter today, $c^{2}_{\rm eff}$, the effective sound speed, and $c^{2}_{\rm vis}$, the viscose sound speed. As already discussed in Section \ref{3rd}, the two last parameters characterize perturbation. In order to derive constraints on the parameters, we used the last version of the MCMC package CosmoMC \citep{Lewis_2002}, that has a convergence diagnostic based on the Gelman and Rubin statistic and includes the support for the Planck data release 2018 Likelihood code \citep{planck2018-V}. 
We assume flat priors on the parameters as listed below in Table \ref{Table1}. For the $\Lambda$CDM parameters, they're significanly wider with respect to the present constraints. For the EDE parameters, we allow for full freedom in the interesting range. 
\begin{table}[!htbp]
\centering
\caption{Flat priors on the cosmological parameters assumed in this paper.}
\label{Table1}
\scalebox{0.8}{
\begin{tabular}{|l|c|}
 \hline 
Parameter & Prior \\
\hline \hline
$\Omega_{\rm b}h^{2}$ & $[0.005,0.1]$  \\
\hline
$\Omega_{\rm c}h^{2}$ & $[0.001,0.99]$  \\
\hline
$100 \theta_{MC}$ & $[0.5,10]$  \\
\hline
$\ln[10^{10}A_{\rm s}]$ & $[1.61,3.91]$ \\
\hline
$n_{s}$ & $[0.8,1.2]$ \\
\hline
$\tau$ & $[0.01,0.8]$ \\
\hline
$\Sigma m_{\nu}\,[{\rm eV}]$ & $[0.056,1]$ \\
\hline
$r$ & $[0,0.3]$ \\
\hline\hline
$\Omega_{\rm eDE}$ & $[0,0.1]$ \\
\hline
$w_{0}$ & $[-1,0]$ \\
\hline
$c^{2}_{\rm eff}$ & $[0,1]$ \\
\hline
$c^{2}_{\rm vis}$ & $[0,1]$ \\
\hline
 \end{tabular} 
}
\end{table}

\subsection{Constraints on EDE}

\begin{figure}[!htbp]
\centering
\includegraphics[width=160mm]{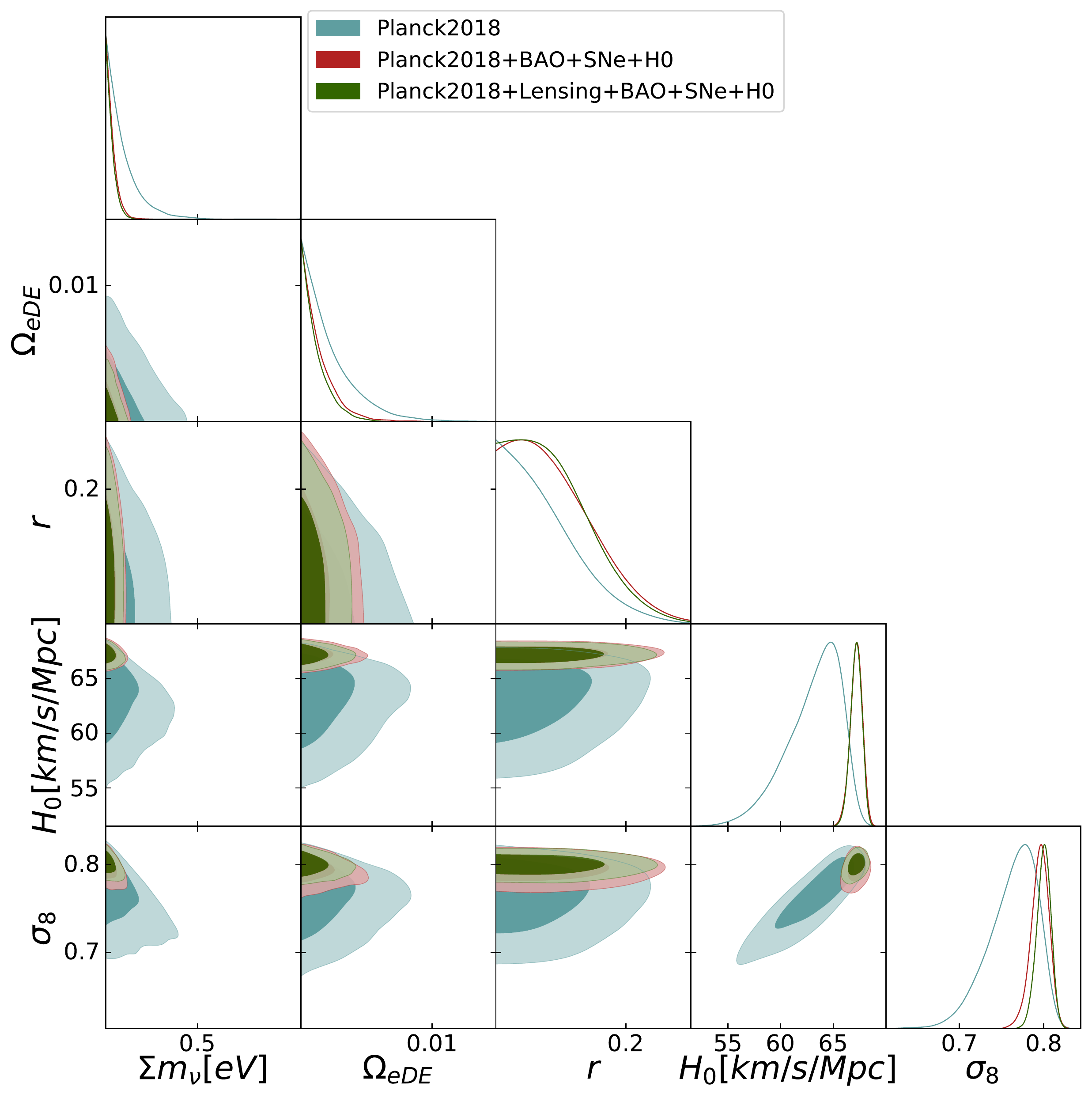}
\caption{Marginalized 2D and 1D posteriors on $\sum m_\nu$, $\Omega_{eDE}$, $r$, $H_0$, $\sigma_8$ including EDE. Blue contours show the 68\% and 95\% confidence level regions allowed from Planck (TT, TE, EE + lowE) \ref{planck} measurements. Contours in red include the BAO \ref{bao}, JLA 
Supernova data set \ref{sne} and $H_{0}$ prior \ref{h0} as well. Green contours, in addition to the previous data set, also include the Planck Lensing \ref{pl-lensing} data. See the first 4 columns of Table \ref{alldata} for the numerical values.}
\label{EDE}
\end{figure}

Table \ref{alldata} and Figure \ref{EDE} show the constraints on cosmological parameters on the EDE scenario.
All parameters, including EDE ones, are allowed to vary within the priors.
Notice that although perturbations are included, both $c_{\rm eff}^2$ and $c_{\rm vis}^2$ turn out to be always unconstrained; therefore we do not show them.
The goal is to see the different constraining power of the combination of data sets, while we will address the role of priors in Section \ref{scenarios}. We can clearly see that by adding BAO, SNe and also the H$_{0}$ prior to the Planck data set, the constraints become tighter (red contours), as expected. By including the lensing data we can see even tighter (green) contour plots, but always overlapping very well with the results by Planck only. The combination of data sets constraining the background and lensing pushes $\Omega_{\rm eDE}$ to lower values, passing from $0.0063$ to $0.0039$ ($\sim40\%$ smaller $2\sigma$ upper limit) and by including the CMB lensing the $2\sigma$ upper limits decreases more to $0.0033$ ($\sim50\%$ smaller than the Planck-only case). The $2\sigma$ upper limit on the parameter $w_0$ becomes much tighter when background data are included, decreasing from $-0.72$ to $-0.95$ (about $\sim 30\%$), while including CMB lensing or QSO does not have a significant impact. 
We can consider the anti-correlation between the equation of state today $w_{0}$ and the tensor to scalar ratio $r$ at least when we are using only Planck data, which is reduced significantly by adding the BAO+SNe+H$_{0}$ data sets. There is also a degeneracy between $\Omega_{\rm eDE}$ and $w_{0}$ and between $\Sigma m_{\nu}$ and the EDE parameters as well. A relative large value of the total neutrino mass $\sim 0.3$ eV would require a small value of $\Omega_{\rm eDE}$, but when background data are included the neutrino mass is much more constrained and somewhat larger values of $\Omega_{\rm eDE}$ can fit the data.

Overall, the main conclusion of this analysis is that the amount of EDE is bound to be well below 1\% at $>2\sigma$ confidence level. This confirms the limits found by the Planck collaboration \citep{planck2015-XIV}, Table 3, where $\Omega_{eDE} < 0.007$ at 2 $\sigma$ CL for fixed neutrino mass. Here we show that these bounds are robust against a variation of the neutrino mass sum, and that they improve once we include lensing and QSO. The degeneracies between the parameters describing the DE model (w$_0$, $\Omega_{\rm eDE}$) and the cosmological parameters that extend the simple vanilla 6-parameter space, neutrino mass and tensor to scalar ratio, are present but are not strong. It is also evident that the Hubble parameter is remarkably stable and constrained to be very close to its $\Lambda$CDM value. On the contrary, $\sigma_8$ inferred from Planck only within this EDE scenario is significantly lower than in $\Lambda$CDM, thus alleviating the tension with the low $\sigma_8$ values inferred from weak lensing \citep{KIDS}. However, the tension is fully restored once background data and CMB lensing are included.

\begin{table}[!htbp]
\centering
\caption{Mean values and $1\sigma$ marginalized error on the cosmological parameters. For $\Sigma m_\nu$, $r$, $\Omega_{\rm eDE}$, and $w_0$ we report the 95\% upper limits. 
}
\label{alldata}
\scalebox{0.7}{
\begin{tabular}{|l|c|c|c|c|}
\hline \hline
Parameter & Planck & Planck+BAO+SNe+$H_0$ & Planck+BAO+SNe+$H_0$+lensing & Planck+BAO+SNe+$H_0$+lensing+QSO \\
\hline
$\Omega_b h^{2}$ &$0.02226 \pm 0.00016 $&$0.02246 \pm 0.00013$&$0.02245 \pm 0.00014$&$0.02245\pm0.00014$\\
$\Omega_c h^{2}$ &$0.1213 \pm 0.0015 $&$0.1187 \pm 0.0010$&$0.1190 \pm 0.0010$&$0.1189\pm0.0010$\\
$100 \theta_{MC}$ &$1.04071\pm0.00034$ &$1.04105\pm0.00030$ &$1.04103\pm0.00030$&$1.04103\pm0.00030$  \\
$\ln(10^{10}A_{s})$ &$3.048\pm0.016$ &$3.048\pm0.016$ &$3.053\pm0.015$&$3.054\pm0.015$ \\
$n_{s}$ &$0.9622\pm0.0048$ &$0.9686\pm0.0040$ &$0.9680\pm0.0039$&$0.9679\pm0.0039$ \\
$\tau$ &$0.0547\pm0.0077$ &$0.0574\pm0.0079$ &$0.0596\pm0.0077$&$0.0597\pm0.0078$ \\
\hline
$\Sigma m_{\nu}$ [eV]  &$<0.31$ &$<0.14$ &$<0.13$&$<0.13$ \\
$r$ &$<0.19$ &$<0.21$ &$<0.20$&$<0.20$ \\
\hline
$\Omega_{eDE}$    &$<0.0063$ &$<0.0039$ &$<0.0033$&$<0.0032$ \\
$w_{0}$           &$<-0.72$ &$<-0.95$ &$<-0.95$&$<-0.96$   \\
\hline
$H_{0}\,[{\rm km/s/Mpc}]$ &$63.13\pm2.60$ &$67.18\pm0.58$ &$67.16\pm0.55$&$67.18\pm0.55$ \\
$\sigma_{8}$ &$0.7632\pm0.0294$ &$0.7954\pm0.0109$ &$0.7999\pm0.0085$&$0.8000\pm0.0085$ \\
\hline \hline

 \end{tabular} 
}
\end{table}

\subsection{Including high redshift expansion tracers}

\begin{figure}[!htbp]
\centering
\includegraphics[width=140mm]{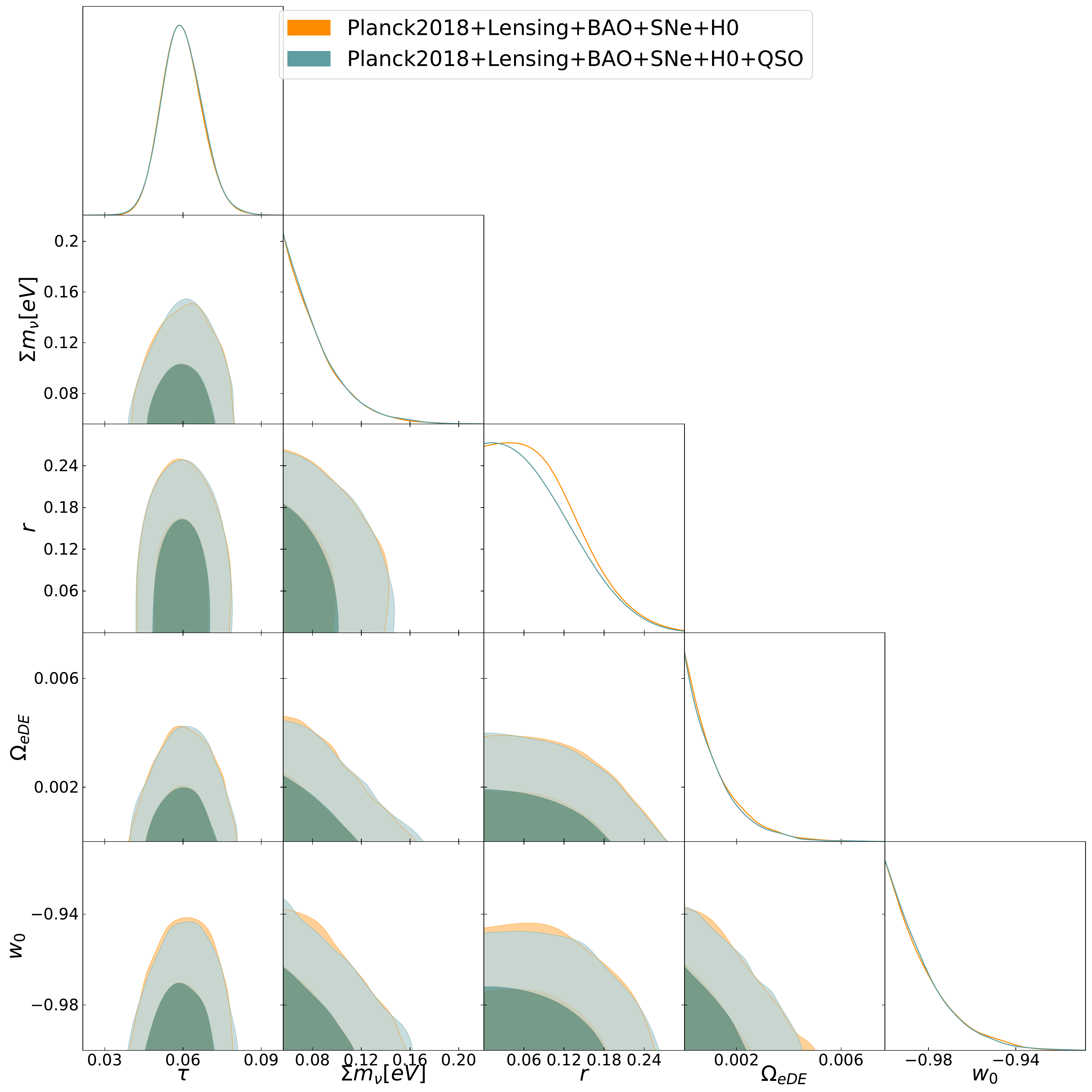}
\caption{Comparison between cosmological constraints within EDE models with and without the QSO data, see the last column of the Table \ref{alldata}}
\label{dr14}
\end{figure}

We will now turn to the question of how much the EDE constraints obtained in the previous Section are affected by the inclusions of high redshift data tracing the cosmological expansion. 
Here we consider the effect of the Hubble diagram of QSOs which we discussed in Section  \ref{qso}. As shown in the last column of Table \ref{alldata} there are no significant differences in the $2 \sigma$ upper limits and also the mean value of the parameters, when QSOs are included. In Figure \ref{dr14} we compare the constraints on cosmological parameters using Planck2018+Lensing+BAO+SNe, with and without the QSO data set. We include the QSO data as well as the other high redshift tracer that we are using in all our analysis, the Lyman-$\alpha$ BAO data. As already mentioned, new BAO data at $z = 2.34$ were obtained from the auto-correlation of Lyman-$\alpha$ forest absorption in eBOSS Data Release 14 \citep[DR14, ][]{Agathe_2019}, as well as from the cross-correlation with quasars in eBOSS DR14 at $z_{\rm eff}=2.35$ \citep{Blomqvist_2019}.    
Therefore present high-redshift tracers of the cosmological expansion do not improve significantly the constraints on the EDE parameters of the EDE model. Similarly, no significant impact is observed for the other parameters. The rationale of including QSOs is that they are a very high redshift probe, and they are useful in particular in addressing their impact on massive neutrinos. Moreover, recent papers found that this quasar dataset prefers a Universe with no dark energy \citep{Yang:2019vgk} (see also \citep{Velten:2019vwo} where the same conclusion is reached in a model-independent way).

\subsection{Cosmological constraints from EDE to $w$CDM}
\label{scenarios}

\begin{figure}[!htbp]
\centering
\includegraphics[width=140mm]{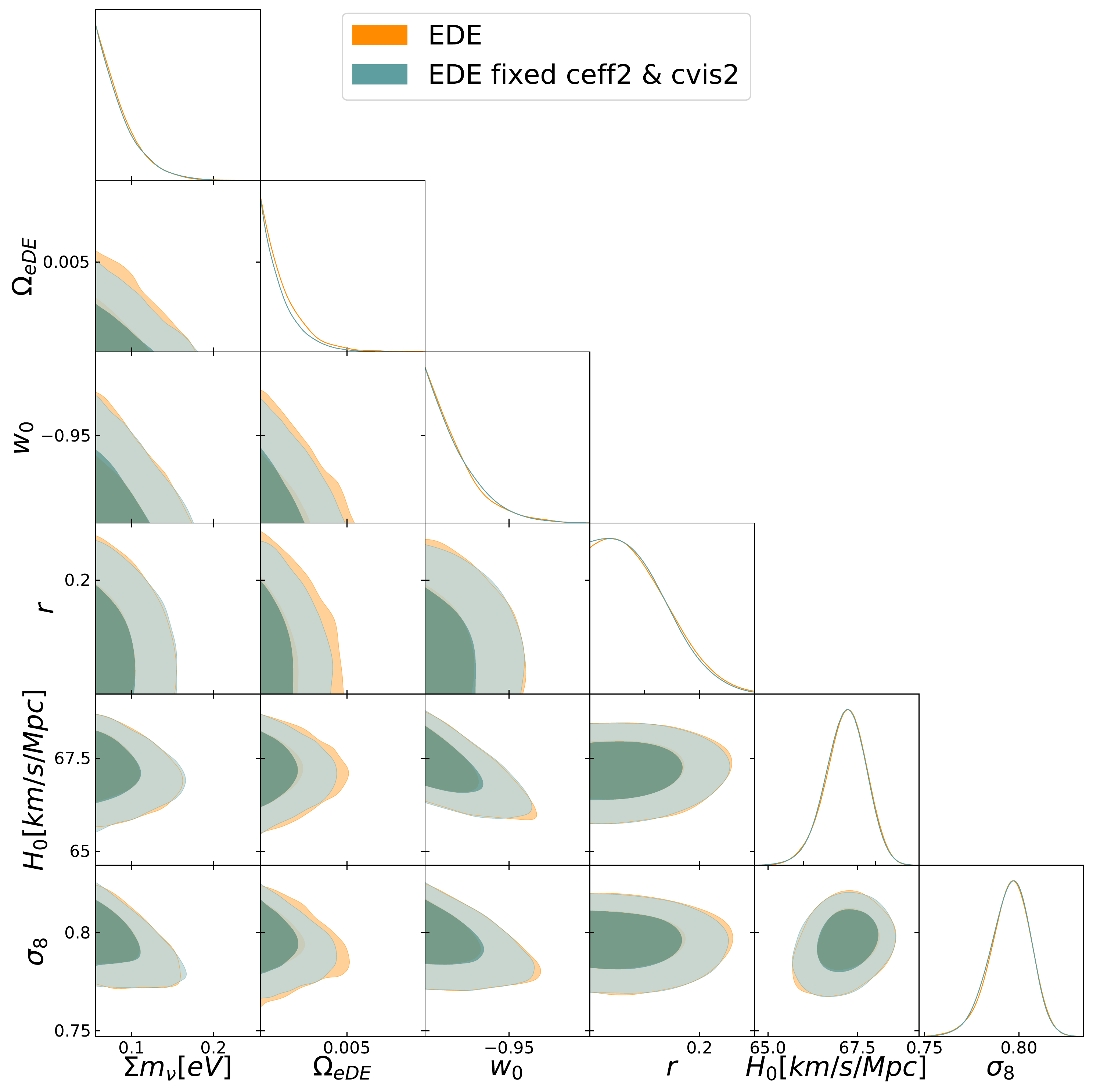}
\caption{Constraints for EDE and EDE fixed $c^{2}_{\rm eff}$ \& $c^{2}_{\rm vis}$ scenarios. Contours show the 68\% and 95\% confidence level regions allowed from Planck+BAO+SNe+$H_0$ measurements.}
\label{blue}
\end{figure}

\begin{figure}[!htbp]
\centering
\includegraphics[width=140mm]{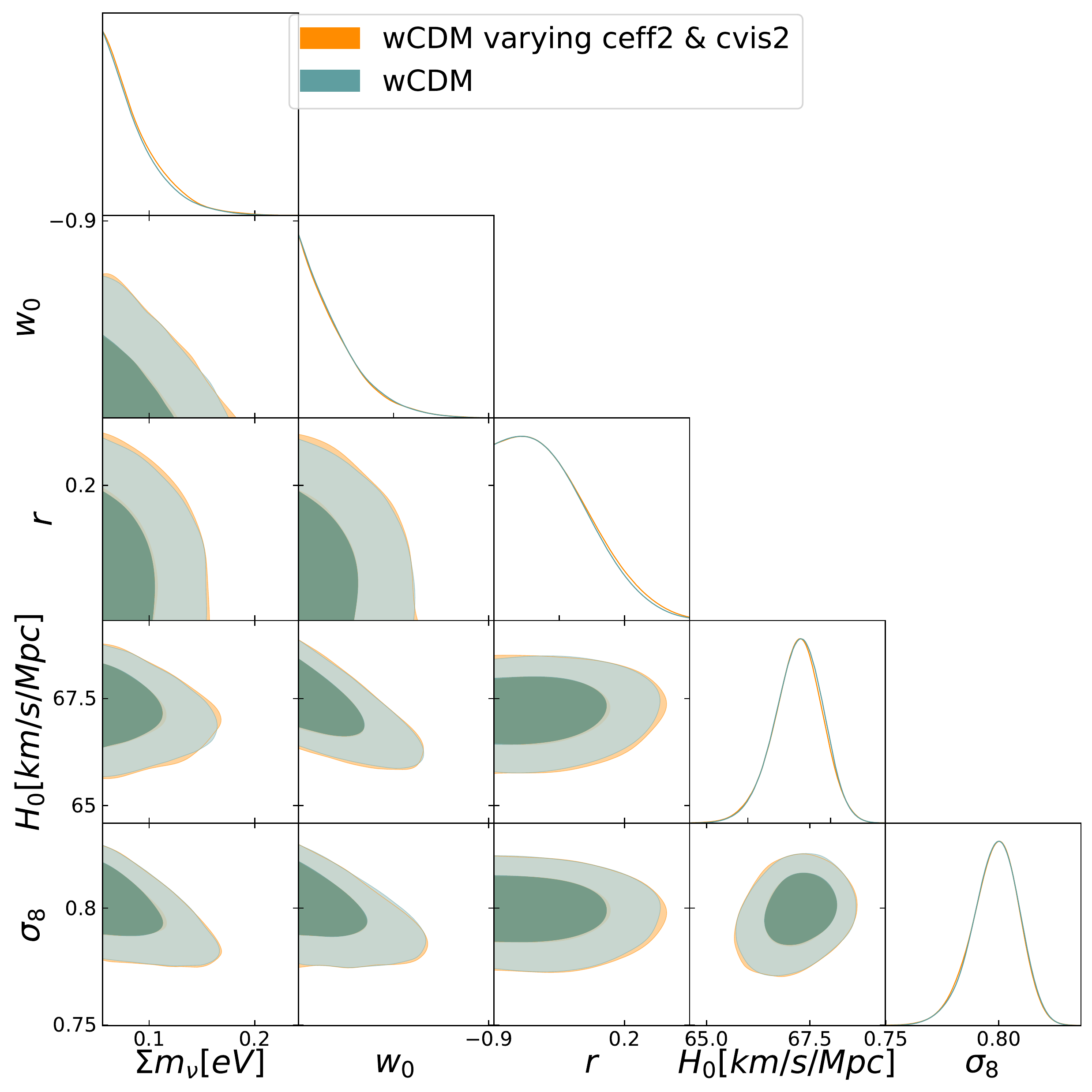}
\caption{"$w$CDM varying $c^{2}_{\rm eff}$ \& $c^{2}_{\rm vis}$" and "$w$CDM" models. Contours show the 68\% and 95\% confidence level regions allowed from Planck+BAO+SNe+$H_0$ measurements.}
\label{red}
\end{figure}

We conclude our analysis by progressively simplifying our EDE models from the general ones to the simple constant equation of state of dark energy, which we refer to as $w$CDM. We make use of priors, listed in Table \ref{Table 7}, while the ones in Table \ref{Table1} are still adopted for the non-EDE cosmological parameters. \\
In the first EDE model, $\Omega_{\rm eDE}$ and $w_{0}$ are allowed to vary, as well as $c^{2}_{\rm eff}$ and $c^{2}_{\rm vis}$. The constraints are shown in Fig. \ref{blue} and listed in the EDE column of Table \ref{EDEtowcdm}. In the second case, named "EDE fixed $c^{2}_{\rm eff}$ \& $c^{2}_{\rm vis}$", perturbations parameters are set to the values $c^{2}_{\rm eff}=1$ and $c^{2}_{\rm vis}=0$, meaning that there is no anisotropic stress in the DE. The constraints are shown in Figure \ref{blue} and Table \ref{EDEtowcdm} - "EDE fixed $c^{2}_{\rm eff}$ \& $c^{2}_{\rm vis}$" column. In a third case, we let no EDE, i.e. $\Omega_{\rm eDE}=0$, while the equation of state today varies as well as the effective and viscosity sound speed. The corresponding constraints are shown in Figure \ref{red} and the fourth column of Table \ref{EDEtowcdm}, named $w$CDM with perturbation. Finally, in the $w$CDM model, only the constant equation of state is allowed to vary, in the range $[-1,0]$. Results are shown in Figure \ref{red} and the last column of Table \ref{EDEtowcdm}. 

\begin{table}[!htbp]
\centering
\caption{Priors for "EDE","EDE fixed $c^{2}_{\rm eff}$ \& $c^{2}_{\rm vis}$", "$w$CDM varying $c^{2}_{\rm eff}$ \& $c^{2}_{\rm vis}$" and "$w$CDM" models.}
\label{Table 7}
\begin{tabular}{|c|c|c|c|c|}
 \hline 
Parameter & EDE & EDE fixed $c^{2}_{\rm eff}$ \& $c^{2}_{\rm vis}$ & $w$CDM varying $c^{2}_{\rm eff}$ \& $c^{2}_{\rm vis}$ & $w$CDM \\
\hline \hline
$\Omega_{\rm eDE}$ & $[0,0.1]$  & $[0,0.1]$ & $0$ &  $0$ \\
\hline
$w_0$ &  $[-1,0]$ & $[-1,0]$ &$[-1,0]$ &$[-1,0]$ \\
\hline
$c_{\rm eff}^{2}$ &  $[0,1]$ & $1$ & $[0,1]$ & $1$ \\
\hline
$c_{\rm vis}^2$ &  $[0,1]$ &  $0$ & $[0,1]$ & $0$ \\
\hline
 \end{tabular} 
\end{table}

\begin{table}[!htbp]
\centering
\caption{Cosmological constraints for Planck2018+BAO+SNe+H0 prior for our 4 different possible scenarios listed in Table \ref{Table 7} .}
\label{EDEtowcdm}
\scalebox{0.8}{
\begin{tabular}{|l|c|c|c|c|}
\hline \hline
Parameter & EDE & EDE fixed $c^{2}_{\rm eff}$ \& $c^{2}_{\rm vis}$ & $w$CDM varying $c^{2}_{\rm eff}$ \& $c^{2}_{\rm vis}$ & $w$CDM \\
\hline
$\Omega_b h^{2}$ &$0.02246 \pm 0.00013$&$0.02246 \pm 0.00014$&$0.02246 \pm 0.00014$& $0.02246\pm0.00014$ \\
$\Omega_c h^{2}$ &$0.1187 \pm 0.0010$&$0.1187 \pm 0.0011$&$0.1186 \pm 0.0011$& $0.1186\pm0.0011$\\
$100 \theta_{MC}$ &$1.04105\pm0.00030$&$1.04102\pm0.00030$ &$1.04107\pm0.00029$ & $1.04107\pm0.00029$ \\
$\ln(10^{10}A_{s})$ &$3.048\pm0.016$ &$3.047\pm0.016$ &$3.047\pm0.016$ & $3.047\pm0.017$ \\
$n_{s}$ &$0.9686\pm0.0040$ &$0.9684\pm0.0040$ &$0.9686\pm0.0041$ & $0.9685\pm0.0041$\\
$\tau$ &$0.0574\pm0.0079$ &$0.0572\pm0.0080$ &$0.0570\pm0.0079$ & $0.0571\pm0.0080$\\
\hline
$\Sigma m_{\nu}\,[{\rm eV}]$ &$<0.14$ &$<0.14$ &$<0.14$& $<0.14$ \\
$r$ &$<0.21$ &$<0.20$ &$<0.21$ &$<0.21$\\
\hline
$\Omega_{eDE}$    &$<0.0039$ &$<0.0034$ &$0$& $0$ \\
$w_{0}$           &$<-0.95$ &$<-0.95$ &$<-0.95$ & $<-0.95$ \\
\hline
$H_{0}\,[{\rm km/s/Mpc}]$ &$67.18\pm0.58$ &$67.16\pm0.58$ &$67.20\pm0.59$ & $67.22\pm0.59$\\
$\sigma_{8}$ &$0.7954\pm0.0109$&$0.7952\pm0.0109$ &$0.7983\pm0.0107$ & $0.7985\pm0.0106$\\
\hline \hline

 \end{tabular} 
}
\end{table}

By looking at results, we derive the following main conclusions. $\Omega_{\rm eDE}$ varies between $0.0034$ to $0.0039$ (2$\sigma$ upper limit), with the tightest limit obtained in the absence of perturbations. The upper limits on the total amount of EDE tend to decrease with an increasing neutrino mass, while with a decreasing value of the tensor to scalar ratio we expect the same limits tend to become less tight.
The parameters describing the DE perturbations are unconstrained and with no significant degeneracy; including them in the analysis does not have any impact on the constraints for the $w$CDM scenario. 
 The bound on the neutrino mass sum is not affected by any of the model extensions shown in Table \ref{EDEtowcdm}.
Finally, with respect to the Planck 2018 fit within $\Lambda$CDM the bounds on the tensor to scalar ratio are relaxed by a factor $2$.


\begin{table}[!htbp]
\centering
\caption{Cosmological constraints for Planck2018+BAO+SNe+H0 prior for the case "EDE fixed $c^{2}_{\rm eff}$ \& $c^{2}_{\rm vis}$" (second column of Table \ref{EDEtowcdm}, corresponding here to the first column), fixing either $\Sigma m_{\nu}$ (second column), or $r$ (third column), or both (fourth column).}
\label{table7}
\scalebox{0.8}{
\begin{tabular}{|l|c|c|c|c|}
\hline \hline
Parameter & EDE fixed $c^{2}_{\rm eff}$ \& $c^{2}_{\rm vis}$ & fixed $\Sigma m_{\nu}=0.056$ & fixed $r=0$ & fixed $\Sigma m_{\nu}=0.056$ and $r=0$ \\
\hline
$\Omega_b h^{2}$ &$0.02246 \pm 0.00014$&$0.02245 \pm 0.00014$&$0.02246 \pm 0.00013$& $0.02246\pm0.00013$ \\
$\Omega_c h^{2}$ &$0.1187 \pm 0.0011$&$0.1189 \pm 0.0010$&$0.1189 \pm 0.0010$& $0.1190\pm0.0010$\\
$100 \theta_{MC}$ &$1.04102\pm0.00030$&$1.04104\pm0.00029$ &$1.04099\pm0.00031$ & $1.04103\pm0.00029$ \\
$\ln(10^{10}A_{s})$ &$3.047\pm0.016$ &$3.047\pm0.016$ &$3.049\pm0.016$ & $3.048\pm0.016$ \\
$n_{s}$ &$0.9684\pm0.0040$ &$0.9683\pm0.0039$ &$0.9669\pm0.0040$ & $0.9666\pm0.0039$\\
$\tau$ &$0.0572\pm0.0080$ &$0.0570\pm0.0079$ &$0.0578\pm0.0078$ & $0.0571\pm0.0079$\\
\hline
$\Sigma m_{\nu}\,[{\rm eV}]$ &$<0.1389$ &$-$ &$<0.1371$& $-$ \\
$r$ &$<0.203$ &$<0.210$ &$-$ &$-$\\
\hline
$\Omega_{eDE}$    &$<0.0034$ &$<0.0035$ &$<0.0036$& $<0.0035$ \\
$w_{0}$           &$<-0.95$ &$<-0.94$ &$<-0.95$ & $<-0.95$ \\
\hline
$H_{0}\,[{\rm km/s/Mpc}]$ &$67.16\pm0.58$ &$67.23\pm0.61$ &$67.13\pm0.58$ & $67.27\pm0.56$\\
$\sigma_{8}$ &$0.7952\pm0.0109$&$0.8003\pm0.0095$ &$0.7966\pm0.0110$ & $0.8009\pm0.0095$\\
\hline \hline

 \end{tabular} 
}
\end{table}

Finally, given that the perturbation parameters are unconstrained, we focus on the case "EDE fixed $c^{2}_{\rm eff}$ \& $c^{2}_{\rm vis}$", and we check whether the varying the neutrino mass and/or the tensor to scalar ratio has any impact on the EDE parameters. The results are shown in Table \ref{table7}. The limits on $\Omega_{eDE}$ are slightly relaxed any time either $\Sigma m_\nu$ or $r$ or both are kept fixed. On the other hand, the bounds on $w_0$ are slightly more loose only if the neutrino mass sum is fixed. Finally, it is interesting to notice that the upper limit on the tensor to scalar ratio in our EDE scenario is quite stable whether the neutrino mass is varying or not. This indicates that the factor 2 in the upper limit on $r$ with respect to $\Lambda$CDM is really induced by EDE.

\section{Conclusions}
\label{conclude}

In light of the present and planned cosmological surveys like DESI\footnote[1]{\label{note1}https://www.desi.lbl.gov} \citep{DESI} / Euclid\footnote[2]{\label{note1} https://sci.esa.int/web/euclid} \citep{EUCLID} / LSST\footnote[3]{\label{note3}https://www.lsst.org} \citep{LSST} and CMB-S4 \footnote[4]{\label{note4}https://cmb-s4.org}\citep{CMB-S4} we are now allowed to reach sub-percent accuracy on the constraints on several cosmological parameters in standard and non-standard scenarios. This can shed light on fundamental physical aspects such as the nature of dark energy, neutrino masses and the physics of inflation.

Here we investigate how cosmological constraints are affected by allowing for an extended framework in the Dark Energy (DE) sector, by considering a perturbed Early Dark Energy (EDE) set of models, involving sound speed and anisotropic stress, implemented in the latest version of the Boltzmann equation solver CAMB \citep{Lewis_2000}. The focus has been on a quantitative exploration of an extended parameter space, considering simultaneous variation of 12 cosmological parameters, using the publicly available MCMC package COSMOMC \citep{Lewis_2002}.

We exploited several of the most important, recent, data sets: the 2018 CMB data from Planck \citep{planck2018-V} in combination with other astrophysical data sets like BAO, Type Ia SNe, the Hubble diagram, Quasar data set and the Lyman-$\alpha$ forest. 

Even by considering this generalized context, we find $\Omega_{\rm eDE} < 0.0039$ at $95\%$ C.L. when Planck, BAO, SNe and the $H_{0}$ prior are considered and in particular EDE perturbations are also included in the modelling. When fixing the latter to default values, we find $\Omega_{\rm eDE} < 0.0034$ at $95\%$ C.L. This quantifies at $\sim15\%$ level the impact of estimating DE perturbation parameters $c^{2}_{\rm eff}$ and $c^{2}_{\rm vis}$ in the fitting. We also find that $c^{2}_{\rm eff}$ and $c^{2}_{\rm vis}$, and their possible departure from standard values, respectively 1 and 0, for a minimally coupled scalar field, are not constrained by present data. 
 
We also checked some cases with fixed values of $\Sigma m_{\nu}$ and $r$ and found that the upper limits on $\Omega_{\rm eDE}$ and $w_{0}$ are remarkably stable even when $\Sigma m_{\nu}$ and the $r$ parameter are varied. Moderate degeneracy is observed between $\Sigma m_{\nu}$ or $r$, and $\Omega_{\rm eDE}$. 

Finally, we also compare the inferred cosmological parameters within the EDE scenario with those obtained in a model with $w$CDM plus perturbations, and with a constant equation of state. We find a non negligible impact on the inferred upper limits on the total neutrino mass and DE equation of state.
Our study confirms the impressive constraining power of the CMB and geometrical probes in setting limits on the evolution of a dark energy component at high redshifts.


\acknowledgments
This research was supported by the COSMOS Network (cosmosnet.it) and Euclid Contract of the Italian Space Agency, as well as by the INDARK INFN Initiative. MV acknowledges financial contribution from the agreement ASI-INAF n.2017-14-H.0.

\bibliographystyle{JHEP}
\bibliography{ref}

\end{document}